\documentclass{aa}

\usepackage{graphicx}
\usepackage{txfonts}
\begin{document}

%#########################################################################################
\title{Global 3D radiation-hydrodynamics models of AGB stars.}
\subtitle{Effects of convection and radial pulsations on atmospheric structures}

\author{B. Freytag
        \and
        S. Liljegren
        \and
        S. H{\"o}fner
}

\institute{Division of Astronomy and Space Physics,
           Department of Physics and Astronomy,
           Uppsala University,
           Box 516,
        SE-751 20 Uppsala,
           Sweden
           (\email{Bernd.Freytag@physics.uu.se})
}

\date{Received ...; accepted ...}

%#########################################################################################
\abstract
% --- Context ---
  {Observations of asymptotic giant branch (AGB) stars
with increasing spatial resolution
reveal new layers of complexity
of atmospheric processes
on a variety of scales.
}
% --- Aim ---
  {To analyze the physical mechanisms that cause asymmetries
and surface structures in observed images,
we use detailed 3D dynamical simulations of AGB stars;
% these stars self-consistently describe convection and pulsations.
these simulations self-consistently describe convection and pulsations.
}
% --- Methods ---
  {We used the CO5BOLD radiation-hydrodynamics code
to produce an exploratory grid of global "star-in-a-box" models
of the outer convective envelope and the inner atmosphere of AGB stars
to study convection, pulsations, and shock waves
and their dependence on stellar and numerical parameters.
}
% --- Results ---
  {The model dynamics are governed by the interaction of
long-lasting giant convection cells,
short-lived surface granules,
and strong, radial, fundamental-mode pulsations.
Radial pulsations and shorter wavelength, traveling, acoustic waves
induce shocks on various scales in the atmosphere.
Convection, waves, and shocks all contribute to the dynamical pressure and, thus, to an increase of the stellar radius
and  to a levitation of material into layers where dust can form.
Consequently, the resulting relation of pulsation period and stellar radius is shifted toward
larger radii compared to that of non-linear 1D models.
The dependence of pulsation period on luminosity agrees well with observed relations.
The interaction of the pulsation mode with the non-stationary convective flow
causes occasional amplitude changes and phase shifts.
The regularity of the pulsations decreases with decreasing gravity
as the relative size of convection cells increases.
The model stars do not have a well-defined surface.
Instead, the light is emitted from a very extended inhomogeneous atmosphere
with a complex dynamic pattern of high-contrast features.
}
% --- Conclusions ---
  {Our models self-consistently describe convection,
convectively generated acoustic noise,
fundamental-mode radial pulsations,
and atmospheric shocks of various scales,
which give rise to complex changing structures in the atmospheres of AGB stars.
}


\maketitle

%#########################################################################################
\section{Introduction}

%% %===============================================================================
%% \subsection{Observations of pulsations}

Variability with typical periods of 100\,--\,1000 days is a
characteristic feature of stars on the asymptotic giant branch (AGB).
The pronounced changes of the stellar luminosity
are generally attributed to low-order large-amplitude pulsations.
Various correlations between pulsation properties and stellar parameters
have been derived from observations, for example, period-luminosity (P-L) relations.
Evolved AGB stars, known as Mira variables,
%\LEt{Depending on your meaning, you could say, "... stars, i.e., Mira variables,..." .}
follow a linear P-L relation,
similar to that of Cepheids,
where a larger luminosity results in a longer period
\citep[see, e.g.,][]{Whitelock2008MNRAS.386..313W, Whitelock2009MNRAS.394..795W}.
Less evolved AGB stars fall on a series of  P-L sequences,
parallel to the sequence of Mira stars
\citep{Wood1999IAUS..191..151W, Wood2015MNRAS.448.3829W}.
These sequences are interpreted as representing stars pulsating in various modes,
with Mira variables pulsating in the fundamental mode
while low-amplitude variables pulsate in overtones.

%% %===============================================================================
%% \subsection{Observed mass loss}

Pulsation and convection seem to play a decisive role in the heavy mass loss
experienced by AGB stars. According to the standard picture, the observed outflows -- with typical
velocities of 5 -- 20 km\,s$^{-1}$ and mass loss rates in the range of
$10^{-8}$ -- $10^{-4} \, {\rm M}_{\sun}$\,yr$^{-1}$ --
are accelerated by radiative pressure on dust grains,
which are formed at about 2\,--\,3 stellar radii.
Pulsations and non-stationary convective flows trigger strong atmospheric shock waves,
which lift gas out to distances
where temperatures are low enough and gas densities are sufficiently high
to allow for the condensation of dust particles
\citep[for a recent review on this mass loss scenario, see, e.g.,][]{Hoefner2015ASPC..497..333H}.

%% %===============================================================================
%% \subsection{Observations of asymmetries and inhomogeneities}

Classically, variable stars have been analyzed via their light curves,
which have been obtained for a large sample of objects
and are usually regular for evolved AGB stars.
However, high-resolution observations of nearby stars
reveal complex irregular structures and dynamical phenomena
on various spatial and temporal scales.

In the case of Mira (o~Cet),
\cite{Karovska1991ApJ...374L..51K} used speckle interferometry with various telescopes
to detect
asymmetries in the extended atmosphere that changed over time.
Later, \cite{Karovska1997ApJ...482L.175K} speculated
that the asymmetries detected with
the Faint Object Camera on the Hubble Space Telescope (HST)
in the UV and optical
might be due to spots or non-radial pulsations.
\cite{Lopez1997ApJ...488..807L} found a best fit to observations in the IR taken with
the Infrared Spatial Interferometer (ISI)
by models with inhomogeneities or clumps.
\cite{Chandler2007ApJ...670.1347C} presented a number of explanations for their ISI observations,
among them shock waves and non-uniform dust formation.
Recently, \cite{Stewart2016MNRAS.457.1410S} produced images of Mira
by analyzing occultations by the rings of Saturn observed with the Cassini spacecraft
showing layers and asymmetries in the stellar atmosphere.

Nevertheless, some phenomena may not be intrinsic to the star itself,
since Mira is part of a binary system and the complex large-scale
structure of the envelope may be attributed to wind-wind interaction
\citep{Ramstedt2014A&A...570L..14R}.

However, similar clumpy and non-spherical structures can be found around other AGB stars.
For instance, very recently, \cite{Ohnaka2016A&A...589A..91O} observed
the semi-regular (type a) star W~Hya
with  VLT/SPHERE-ZIMPOL and VLTI/AMBER and find
supporting evidence for clumpy dust clouds caused by
pulsations and large convective cells,
as predicted by 3D simulations for AGB stars
\citep{Freytag2008A&A...483..571F}.

A large number of observations
% \LEt{Please see listing for "quite" in Sect. 8 of the Author's Guide
% (www.aanda.org/language-editing) and check for this throughout.}
exist of the carbon star IRC+10216 (CW~Leo),
starting with the first detection of a bipolar asymmetry
in coronographic images
by \cite{Kastner1994ApJ...434..719K}.
\cite{Haniff1998A&A...334L...5H}
attribute asymmetries in diffraction-limited interferometric images
to envelope clearing along a bipolar axis.
\cite{Weigelt1998A&A...333L..51W} used speckle-masking interferometry with the SAO 6\,m telescope
% to find an ``extremely clumpy'' dust shell, that
% ``gives most likely direct evidence for an inhomogeneous mass-loss
% process which may be interpreted in terms of large-scale
% surface convection-cells \citep{Schwarzschild1975ApJ...195..137S}
% being a common phenomenon for red giants.''
% \LEt{Please avoid using a quotation like this in which there is a reference within the quotation.
%      Please paraphrase and cite using (Weigelt et al. 1998 and references therein).}
to find a clumpy dust shell, that
was considered as direct evidence for an inhomogeneous mass-loss process
caused large-scale surface convection-cells,
whose presence was suggested by \cite{Schwarzschild1975ApJ...195..137S}.
Recently, \cite{Stewart2016MNRAS.455.3102S} investigated the dynamical evolution of dust clouds
in images reconstructed from aperture-masking interferometric observations using the Keck and VLT
and from occultation measurements by Cassini,
resulting in a complete
change in the asymmetry and distribution of the clumps
compared to the observations about 20\,years earlier by
\cite{Kastner1994ApJ...434..719K,
Haniff1998A&A...334L...5H,
Weigelt1998A&A...333L..51W}.

The reason for these observed asymmetries could be the
underlying irregular global convective flows in AGB stars,
although the star itself or the stellar surface are not directly visible.

%% %===============================================================================
%% \subsection{1D models}

A realistic modeling of the underlying physical processes is a
notoriously difficult problem
and theoretical studies found in the literature are usually
restricted to 1D simulations.
Still, dynamical 1D atmosphere and wind models
have been used successfully to explore the basic
dust-driven mass-loss process,
relying on a parameterized description of
sub-photospheric velocities due to radial pulsations
\citep[so-called piston models; see, e.g.,][]{Bowen1988ApJ...329..299B,
Fleischer1992A&A...266..321F,
Winters2000A&A...361..641W,
Hoefner2003A&A...399..589H,
Jeong2003A&A...407..191J,
Hoefner2008A&A...491L...1H}.
Mass loss rates and wind velocities along with spectral energy distributions,
variations of photometric colors with pulsation phase,
and other observable properties, resulting from the latest generation
of such time-dependent atmosphere and wind models computed with the DARWIN code,
show good agreement with observations
\citep{Nowotny2010A&A...514A..35N,
Eriksson2014A&A...566A..95E,
Bladh2015A&A...575A.105B,
Hoefner2016A&A...594A.108H}.

One-dimensional CODEX models of gas dynamics in the stellar interior and atmosphere,
simulating (radial) Mira-type pulsations as self-excited oscillations
and following the propagation of the resulting shock waves
through the stellar atmosphere,
have been presented by
\cite{Ireland2008MNRAS.391.1994I,Ireland2011MNRAS.418..114I}, for instance.

%% %===============================================================================
%% \subsection{3D models}

Despite their success in reproducing various observational results,
1D dynamical models are not sufficient to give a comprehensive
picture of the physical processes leading to mass loss on the AGB.
These 1D models require a number of free parameters
that have to be carefully adjusted,
for example, to describe the variable inner boundary
("piston") in dynamical atmosphere and wind models
or the time-dependent extension of the mixing-length theory (MLT)
in non-linear pulsation models.
%Further, such models cannot intrinsically simulate 3D
Further, such models cannot simulate intrinsically 3D
phenomena such as stellar convection
and they therefore cannot describe the giant convection cells that are
presumably giving rise to the non-spherical and clumpy morphology of the atmosphere.

Turbulent flows, in general, and stellar convection, in particular,
are known to produce acoustic waves if the Mach number
is sufficiently large \citep{Lighthill1952RSPSA.211..564L}.
This excitation process has been studied with local 3D radiation-hydrodynamics
simulations for the case of the Sun, for example, by
\cite{Nordlund2001ApJ...546..576N} and \cite{Stein2001ApJ...546..585S}.
The steepening of these waves while traveling upward
into the chromosphere was investigated by
\cite{Wedemeyer2004A&A...414.1121W}.
Such local 3D radiation-hydrodynamics simulations have been used for decades to
model small patches on the surface of (more or less) solar-type
stars. Now a number of grids
produced by different groups with various codes
are available
\citep{Ludwig2009MmSAI..80..711L,
Magic2013A&A...557A..26M,
Trampedach2013ApJ...769...18T,
Beeck2013A&A...558A..48B},
which also extend to the regime of white dwarfs
\citep{Tremblay2015ApJ...799..142T}.

In contrast to these compact types of stars,
giants might be covered by only a small number of huge convective cells,
as suggested by \cite{Stothers1971A&A....10..290S}
as an explanation for irregularities in the light curves.
\cite{Schwarzschild1975ApJ...195..137S} argued that,
if the size of surface convection cells is governed by some characteristic length scale, such as the density scale height,
the counterpart of small solar granules should be huge convective structures on giants.
Consequently, these stars should produce large-scale acoustic waves.
Surface convection and associated waves, pulsation, and shocks
have been investigated with global 3D radiation-hydrodynamics simulations
with the CO5BOLD code \citep{Freytag2012JCP...231..919F},
both for the case of red supergiants,
for instance, by \cite{Freytag2002AN....323..213F} and \cite{Chiavassa2009A&A...506.1351C},
and with an exploratory study for an AGB star by \cite{Freytag2008A&A...483..571F}.
These studies confirm the existence of large-scale convection cells and acoustic waves.

In this paper, we present the first grid of AGB star models
produced with a new, improved version of CO5BOLD.
We analyze the dependence of the properties of convection and pulsations on stellar parameters
and also look at the influence of rotation.

%...............................................................................
% --- Table of model parameters, produced with rhd_printtable.pro ---
\begin{table*}[htb]
 \begin{center}
  \caption{Basic model parameters
  \label{t:ModelParam}}
  \begin{tabular}{l|rrrrrr|rrrrrr}
\hline
model & $M_\star$ & $M_\mathrm{env}$ & $L_\star$ &\!\!$n_x$$\times$$n_y$$\times$$n_z$ & $x_\mathrm{box}$ & $P_\mathrm{rot}$ & $t_\mathrm{avg}$ & $R_\star$ & $T_\mathrm{eff}$ & $\log g$ & $P_\mathrm{puls}$ & $\sigma_\mathrm{puls}$ \\
 & $M_\sun$ & $M_\sun$ & $L_\sun$ &  & $R_\sun$ & yr & yr & $R_\sun$ & K & (cgs) & yr & yr \\ \hline
 st28gm06n02 & 1.0 & 0.196 &  7079 & 127$^3$ & 1244 &$\infty$& 11.01 &  437 & 2531 & -0.85 & 1.400 & 0.956 \\
 st28gm06n03 & 1.0 & 0.188 &  6589 & 171$^3$ & 1674 &$\infty$&  2.41 &  400 & 2599 & -0.77 & 1.379 & 0.351 \\
 st28gm06n05 & 1.0 & 0.187 &  8144 & 171$^3$ & 1674 &$\infty$&  2.06 &  423 & 2664 & -0.82 & 1.775 & 1.093 \\
 st28gm06n06 & 1.0 & 0.186 &  6905 & 171$^3$ & 1674 &$\infty$&  4.48 &  430 & 2538 & -0.83 & 1.420 & 0.907 \\ \hline
 st28gm06n13 & 1.0 & 0.181 &  6932 & 281$^3$ & 1381 &$\infty$& 29.96 &  384 & 2687 & -0.73 & 1.479 & 0.586 \\
 st28gm06n16 & 1.0 & 0.178 &  6582 & 401$^3$ & 1381 &$\infty$& 23.30 &  395 & 2616 & -0.76 & 1.376 & 0.449 \\
 st28gm06n18 & 1.0 & 0.182 &  6781 & 401$^3$ & 1970 &$\infty$& 26.75 &  395 & 2635 & -0.76 & 1.325 & 0.459 \\
 st28gm06n24 & 1.0 & 0.182 &  6944 & 281$^3$ & 1381 &$\infty$& 23.77 &  372 & 2733 & -0.71 & 1.262 & 0.339 \\
 st28gm06n25 & 1.0 & 0.182 &  6890 & 401$^3$ & 1970 &$\infty$& 23.77 &  372 & 2727 & -0.71 & 1.388 & 0.360 \\
 st28gm06n29 & 1.0 & 0.182 &  6956 & 281$^3$ & 1381 &     20 & 25.35 &  384 & 2688 & -0.73 & 1.297 & 0.337 \\
 st28gm06n30 & 1.0 & 0.182 &  6951 & 281$^3$ & 1381 &     10 & 25.34 &  395 & 2652 & -0.76 & 1.327 & 0.200 \\ \hline
st28gm07n001 & 1.0 & 0.176 & 10028 & 281$^3$ & 1381 &$\infty$& 30.90 &  531 & 2506 & -1.02 & 2.247 & 1.397 \\
st26gm07n002 & 1.0 & 0.544 &  6986 & 281$^3$ & 1381 &$\infty$& 25.35 &  437 & 2524 & -0.85 & 1.625 & 0.307 \\
st26gm07n001 & 1.0 & 0.315 &  6953 & 281$^3$ & 1381 &$\infty$& 27.74 &  400 & 2635 & -0.77 & 1.416 & 0.256 \\
 st28gm06n26 & 1.0 & 0.182 &  6955 & 281$^3$ & 1381 &$\infty$& 25.35 &  371 & 2737 & -0.70 & 1.290 & 0.317 \\
st29gm06n001 & 1.0 & 0.109 &  6948 & 281$^3$ & 1381 &$\infty$& 25.35 &  348 & 2822 & -0.65 & 1.150 & 0.314 \\
st27gm06n001 & 1.0 & 0.548 &  4982 & 281$^3$ & 1381 &$\infty$& 28.53 &  345 & 2610 & -0.64 & 1.230 & 0.088 \\
st28gm05n002 & 1.0 & 0.262 &  4978 & 281$^3$ & 1381 &$\infty$& 25.35 &  313 & 2742 & -0.56 & 1.077 & 0.104 \\
st28gm05n001 & 1.0 & 0.182 &  4990 & 281$^3$ & 1381 &$\infty$& 25.36 &  300 & 2798 & -0.52 & 1.026 & 0.135 \\
st29gm04n001 & 1.0 & 0.141 &  4982 & 281$^3$ & 1381 &$\infty$& 25.35 &  294 & 2827 & -0.50 & 0.927 & 0.100 \\
\hline
  \end{tabular}
 \end{center}
{The table shows
the model name, composed of the approximate effective temperature and surface gravity
  and of a running number;
the mass $M_\star$, used for the external potential;
the envelope mass $M_\mathrm{env}$, derived from integrating the mass density of all
grid cells within the computational box;
the average emitted luminosity $L_\star$, close but not identical to the inserted luminosity of either
  5000, 7000, or 10000\,$L_\sun$ in the core;
the model dimensions $n_x$$\times$$n_y$$\times$$n_z$;
the edge length of the cubical computational box $x_\mathrm{box}$;
the rotational period $P_\mathrm{rot}$; the time $t_\mathrm{avg}$, used for the averaging of the rest of the quantities in this table and for the further analysis;
the average approximate stellar radius $R_\star$;
the average approximate effective temperature $T_\mathrm{eff}$;
the logarithm of the average approximate surface gravity $\log g$;
the pulsation period $P_\mathrm{puls}$;
and the half width of the distribution of the pulsation frequencies $\sigma_\mathrm{puls}$.
The first (``old'') group of simulations comprises models used in \cite{Freytag2008A&A...483..571F}.
In the second (``test'') group, numerical parameters (e.g., the box size or the number of grid points)
  or the rotation period were varied.
The last (``grid'') group comprises models with slightly varying stellar parameters
($M_\mathrm{env}$,  $L_\star$).
}
\end{table*}
%...............................................................................

%...............................................................................
\begin{figure}
\centering
\includegraphics[width=8.8cm]{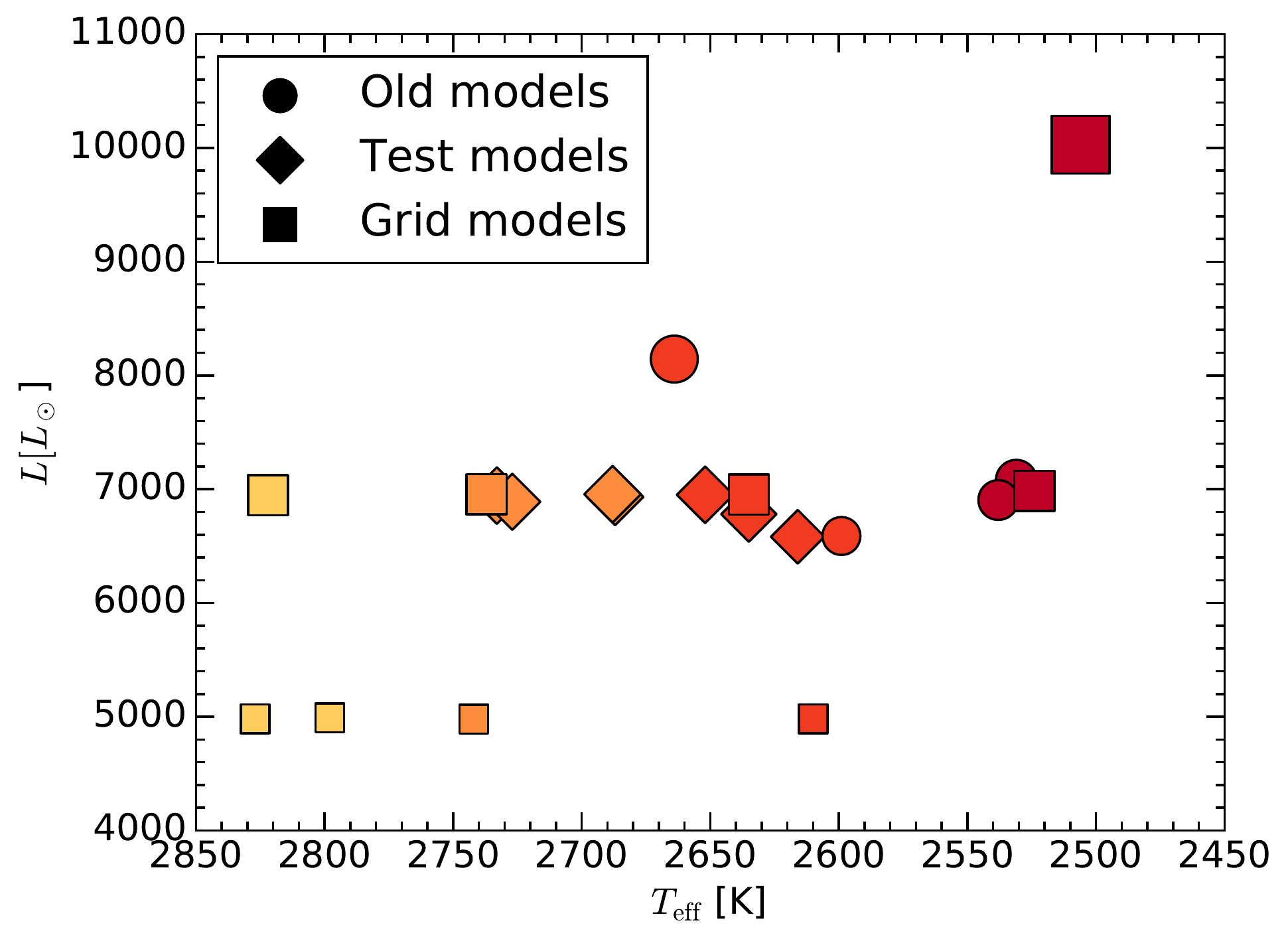}
\caption{Luminosity and effective temperature of all models.
The colors of the markers represent the temperature of the models
while the size of the markers represent the luminosity.
The shape of the markers indicates which group in Table \ref{t:ModelParam}
the models belong to;
the ``old'' models from \citep{Freytag2008A&A...483..571F} are circles,
the ``test'' models with varied numerical parameters are diamonds, and
the ``grid'' models with different stellar parameters are squares.}
\label{f:overview_models}
\end{figure}
%...............................................................................

%#########################################################################################
\section{Setup of global AGB star models with CO5BOLD}\label{s:Setup}

Following our earlier work \citep{Freytag2008A&A...483..571F},
we present new radiation-hydrodynamics simulations of AGB stars with CO5BOLD
\citep{Freytag2012JCP...231..919F}.
Improvements regarding the accuracy and stability of
the hydrodynamics solver are outlined in
\cite{Freytag2013MSAIS..24...26F}.

The CO5BOLD code
% \LEt{Please avoid beginning sentences with abbreviations, acronyms, numbers in figures, and the like.
% Please check for this throughout the paper.  See Sect. 6.2.1.}
solves the coupled non-linear equations of
compressible hydrodynamics (with an approximate Roe solver)
and non-local radiative energy transfer
(for global models with a short-characteristics scheme)
in the presence of a fixed external gravitational field.
The numerical grid is Cartesian.
In all models presented here, the computational domain and all individual
grid cells are cubical.
The tabulated equation of state takes the ionization of hydrogen
and helium and the formation of H$_2$ molecules into account.
It is assumed that solar abundances are appropriate for M-type AGB stars.
The tabulated gray opacities
\citep[very similar to those used in][]{Freytag2008A&A...483..571F}
are merged from Phoenix \citep{Hauschildt1997ApJ...483..390H}
and OPAL \citep{Iglesias1992ApJ...397..717I} data at around 12\,000\,K,
with a slight reduction in the very cool layers (that are hardly
reached in the current model grid) to remove any influence of dust
onto the opacities.
There are no source terms or dedicated opacities for dust:
no dust is included in any of the current models
in contrast to the two old models st28gm06n05 and n06.

The gravitational potential is spherically symmetric,
\citep[see Eq.\,(41) in][]{Freytag2012JCP...231..919F}
corresponding to a 1\,$M_\sun$ core in the outer layers and smoothed in the center
at $r$\,$\lesssim$\,$R_0$\,=\,78\,$R_\sun$
\citep[see Fig.\,4 in][]{Freytag2008A&A...483..571F}.
The tiny central nuclear-reaction region cannot possibly be resolved with grid cells
of constant size.
Instead, in the smoothed-core region,
a source term feeds in energy corresponding to
5000, 7000, or 10000\,$L_\sun$.
A drag force is only active in this core to prevent dipolar flows
traversing the entire star.
All outer boundaries are open for the flow of matter and for radiation.

The start models for the first AGB simulations
presented in \cite{Freytag2008A&A...483..571F}
were based on hydrostatic 1D stratifications
\citep[see Fig.\,1 of][showing images from the initial phase of a
simulation of a red supergiant started this way]{Chiavassa2011A&A...535A..22C}.
All later AGB models were derived from a predecessor 3D model by
either refining or coarsening the numerical grid,
adding or removing grid layers to change the size of the computational box,
increasing or decreasing the core luminosity in time in steps of 1000\,$L_\sun$,
or by modifying the density in each grid cells in each time step by a tiny amount
to change the initial to the final envelope mass.
While the initial adjustment time lasted typically a few years,
the typical total simulation time span is 10$^9$\,s (just above 30\,yr).
We monitored the time evolution of several quantities, for instance, the total emerging luminosity and the mass density in the very center,
to ensure that the models are well relaxed when we compute averaged quantities
for further analysis (see below).

Table\,\ref{t:ModelParam} and Fig.\,\ref{f:overview_models} give
an overview of the simulations split into three groups:
old models from \cite{Freytag2008A&A...483..571F},
new test models used for parameter studies (including rotation rate),
and a grid of models with different stellar parameters.

While, for instance, the mass $M_\star$ (controlling the gravitational potential),
the resolution and the extent of the numerical grid, and
the rotation rate are pre-chosen fixed parameters
(second group of rows in Table\,\ref{t:ModelParam}),
other model properties are determined after a simulation is finished
(third group in Table\,\ref{t:ModelParam}).
The stated stellar luminosity is a time average of the luminosity for each
``fine'' snapshot without the full 3D information, but
with lots of preprocessed -- and thus compressed -- data,
saved every 250\,000\,s\,$\approx\,$3\,d.
The envelope mass $M_\mathrm{env}$, however, is calculated from the integrated density for
each fine snapshot averaged over time.
Nevertheless, the radius is more difficult to determine and less well defined.
The radius is chosen as that point $R_\star$
where the spherically (abbreviated as $\langle . \rangle_\Omega$)
and temporally (denoted as $\langle . \rangle_t$) averaged temperature
and the average luminosity $\langle L \rangle_{\Omega,t}$
fulfill $\langle L \rangle_{\Omega,t}$\,=$\,4\pi\sigma R_\star^2 \langle T \rangle_{\Omega,t}^4$.
Then, effective temperature and surface gravity follow.

To investigate purely radial motions we take averages over spherical shells for each fine snapshot
for the radial mass flux $\langle \rho v_\mathrm{radial} \rangle_\Omega(r,t)$
and the mass density  $\langle \rho \rangle_\Omega(r,t)$
and take the ratio as the radial velocity, which is now a function
$\langle v_\mathrm{radial} \rangle(r,t)$
of radial distance and time.
The derivation of the pulsation quantities $P_\mathrm{puls}$ and $\sigma_\mathrm{puls}$
of Table\,\ref{t:ModelParam}
is described in Sect.\,\ref{s:Pulsations}.

%...............................................................................
% --- aaagb2_intseq_plot.pro ---
\begin{figure*}[hbtp]
\begin{center}
%\hspace*{0.8cm}\includegraphics[width=3.44cm]{aaagb2_st28gm06n25_T_mean.pdf}
%\includegraphics[width=3.44cm]{aaagb2_st28gm06n25_rho_mean.pdf}
%\includegraphics[width=3.44cm]{aaagb2_st28gm06n25_s_mean.pdf}
%\includegraphics[width=3.44cm]{aaagb2_st28gm06n25_s_str_mean.pdf}
%\includegraphics[width=3.44cm]{aaagb2_st28gm06n25_Int3l_mean.pdf}\vspace{0.2cm}
\includegraphics[width=4.275cm]{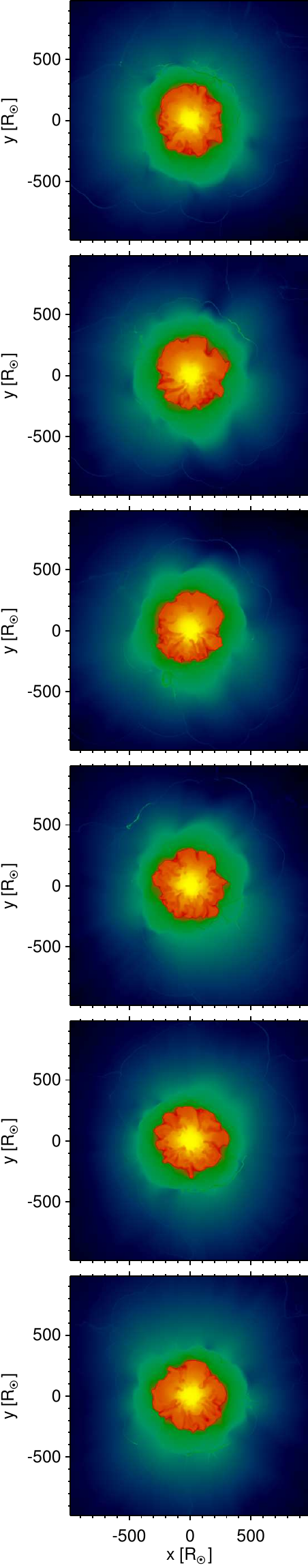}
\includegraphics[width=3.42cm]{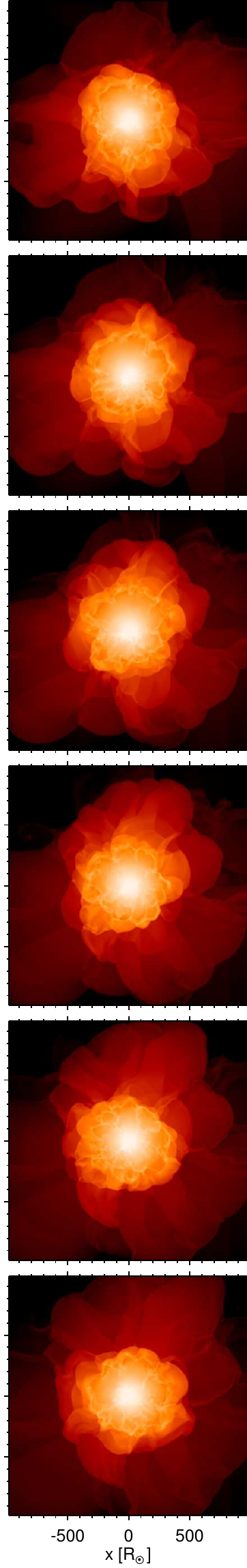}
\includegraphics[width=3.42cm]{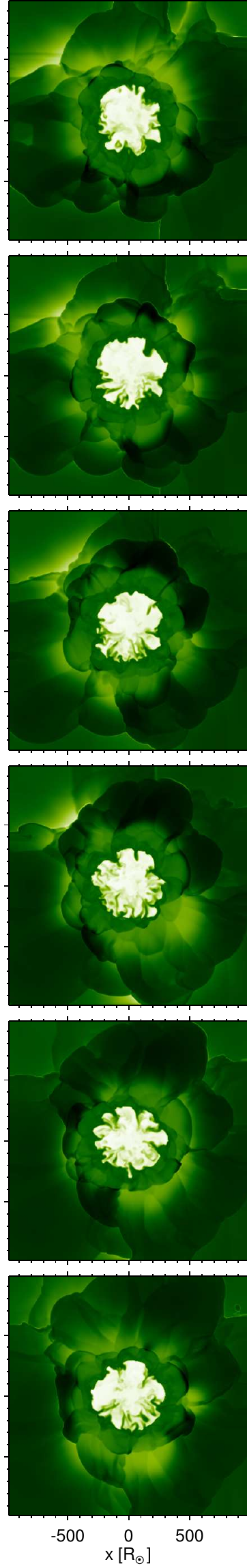}
\includegraphics[width=3.42cm]{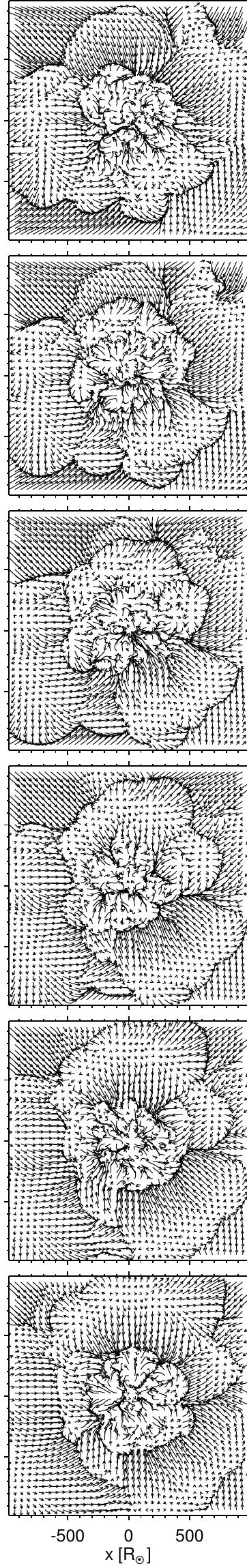}
\includegraphics[width=3.42cm]{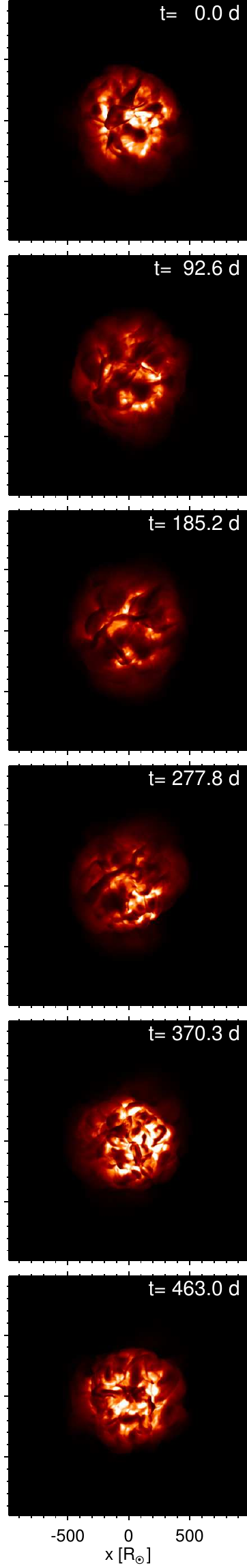}
\end{center}
\caption{Time sequences (for the extended model st28gm06n25) of
temperature slices,
density slices,
entropy slices,
pseudo-streamlines (integrated over about 3 months),
and bolometric intensity maps,
where black corresponds to zero intensity.
The snapshots are nearly 3 months apart
(see the counter in the right-most panels),
so that the entire sequence covers about one pulsational cycle.
The color scales are kept the same for all snapshots.
This figure can be directly compared to Fig.\,1 of \cite{Freytag2008A&A...483..571F}.}
%\vspace{0.9cm}
\label{f:st28gm06n25_QuSeq}
\end{figure*}
%...............................................................................

%#########################################################################################
\section{Results}

%===============================================================================
\subsection{General dynamics and comparison to old models}

\begin{figure*}[hbtp]
% \hspace*{0.9cm}\includegraphics[width=17.3cm]{aaagb2_st28gm06n26_Int3r_mean.pdf}\vspace{0.4cm}
\includegraphics[width=17.8cm]{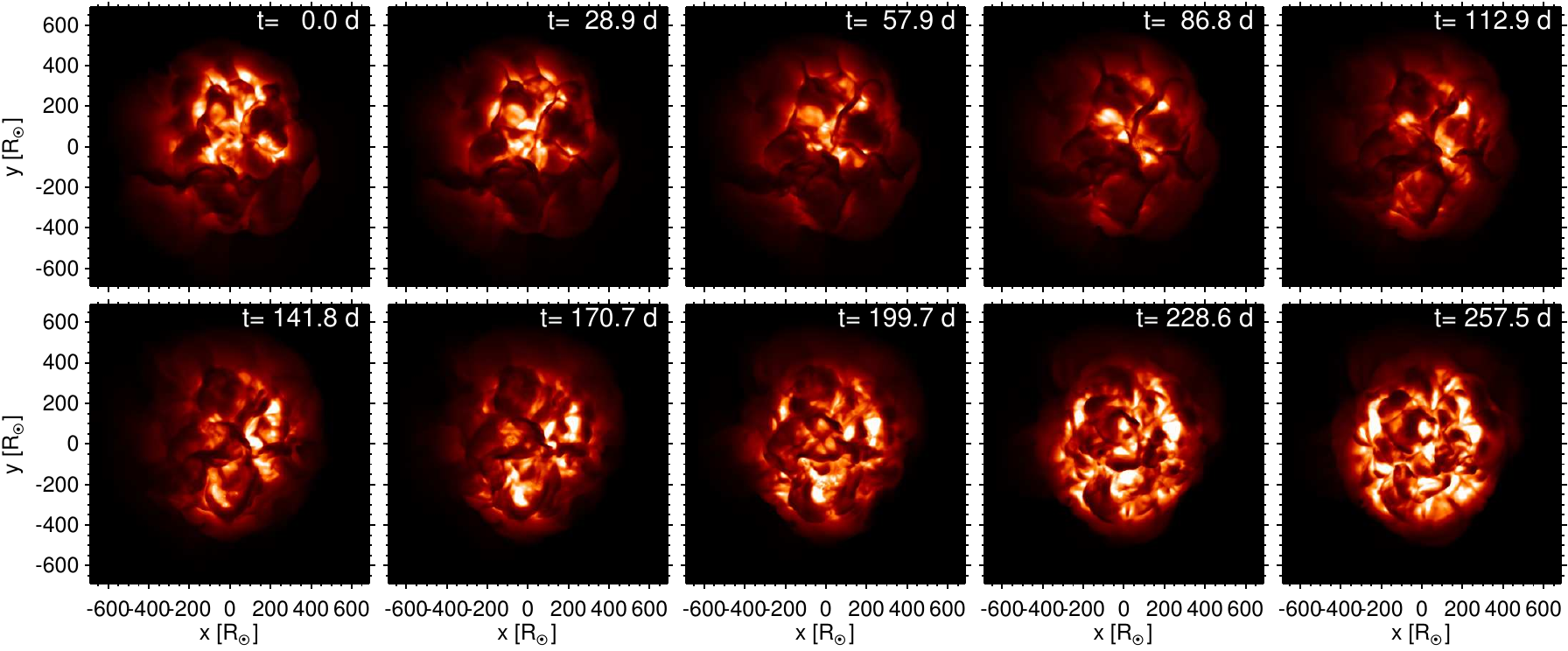}
\caption{Time sequence (for the standard model st28gm06n26) of
bolometric intensity maps,
where black corresponds to zero intensity.
The snapshots are about 1~month apart to demonstrate the relative short timescale
for changes of the small surface features.}
\label{f:st28gm06n26_IntSeq}
\end{figure*}

\begin{figure*}
\centering
\includegraphics[width=18cm]{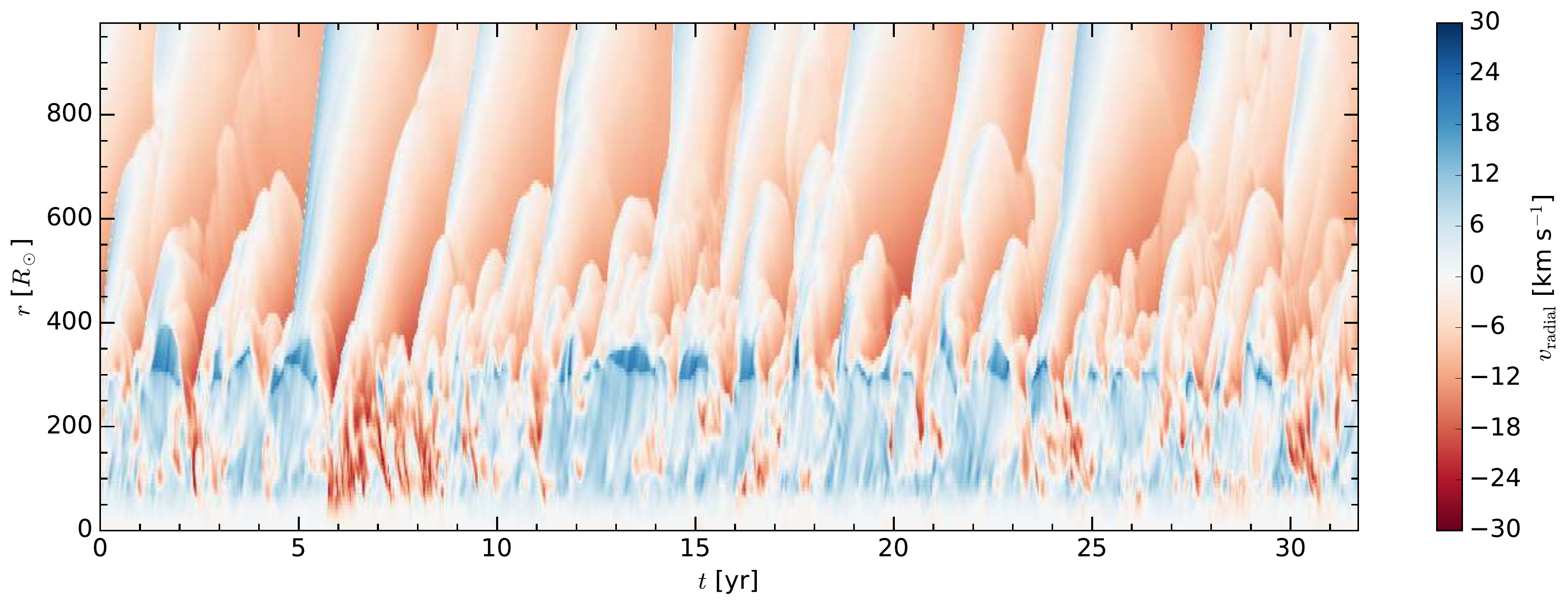}
\caption{Plot (for the standard model st28gm06n26)
of radial velocity $v_\mathrm{radial}(r,t)$ for all grid points in one column
from the center to the side of the computational box
for the entire simulation time.
Blue indicates outward and red inward flow.}
\label{f:vel_field_1}
\end{figure*}

\begin{figure*}
\centering
\includegraphics[width=18cm]{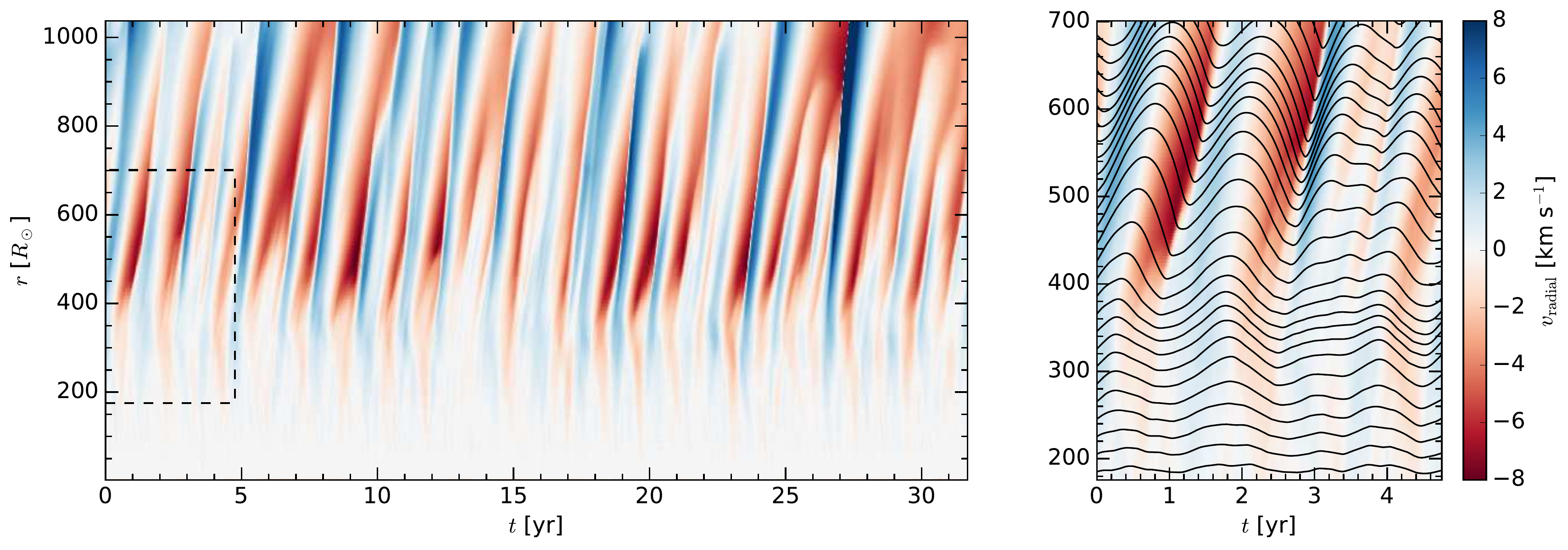}
\caption{Examples for the standard model st28gm06n26:
% Representation of the standard model st28gm06n26.
% \LEt{Please fix according to intended meaning, if necessary.}
\textit{Left}: The spherically averaged radial velocity $\langle v_\mathrm{radial}\rangle(r,t)$
for the full run time and radial distance.
The different colors show the average vertical velocity at that time and radial distance.
\textit{Right:} Part of the velocity field from the right image,
indicated with the rectangle with mass-shell movements plotted as iso-mass contour lines.}
\label{f:vel_field}
\end{figure*}

The time evolution of various quantities can be followed in the
snapshot sequences in Figs.\,\ref{f:st28gm06n25_QuSeq} and \ref{f:st28gm06n26_IntSeq};
the plots of radial velocity as function of radius and time
are shown in Fig.\,\ref{f:vel_field_1} (along a particular radius vector: $v_\mathrm{radial}(r,t)$)
and Fig.\,\ref{f:vel_field} (averaged over spheres: $\langle v_\mathrm{radial}\rangle(r,t)$).

%...............................................................................
% --- aaagb2_intseq_plot.pro ---
%...............................................................................

%...............................................................................
%...............................................................................

%...............................................................................
%...............................................................................

%===============================================================================
\subsection{Convective motions}\label{s:Convection}

% The convection zone is essentially shown with the bright (high entropy) irregular inner part
The convection zone is essentially indicated by the bright (high entropy) irregular inner part
of the entropy snapshots in Fig.\,\ref{f:st28gm06n25_QuSeq} with
a radius around 300\,$R_\sun$, which is smaller than the inferred stellar radius
of $R_\star$\,=\,371\,$R_\sun$
given in Table\,\ref{t:ModelParam};
i.e., light is emitted from layers further out.
The drop in entropy at the top of the convective layers is accompanied
by a drop in temperature and even a thin density-inversion layer.

Huge convective cells can span 90\,degrees or more in the cross sections, which is
in line with extrapolations by \cite{Schwarzschild1975ApJ...195..137S}
from solar granulation.
The cells are outlined by non-stationary downdrafts
reaching from the surface of the convection zone
to the center of the model star.
While the flow travel time from the surface to the center
is around half a year,
the convective cells can have a lifetime of many years,
causing long intervals of one preferred flow direction in the convection zone
in Fig.\,\ref{f:vel_field_1}.
These downdrafts even have a tendency to traverse the artificial stellar core
and to create a global dipolar flow field.
However, in the (non-rotating) models presented here,
the drag force, which is somewhat arbitrarily applied in the stellar core,
prevents these flows.

The surface of the convection zone appears corrugated;
this is caused by many smaller short-lived convection cells close to the surface.
% this is caused by many smaller short-lived ``granulation'' cells close to the surface.
% \LEt{Please avoid the use of quotation marks unless you are actually quoting
% something. Please also avoid the use of italics used for emphasis.
% See Sect. 1.2 of the Author's Guide. Please check for these throughout.}
Their number increases with numerical resolution,
as a comparison of Fig.\,\ref{f:st28gm06n25_QuSeq}
with Fig.\,1 of \cite{Freytag2008A&A...483..571F} reveals.

Simulations of red supergiants (RSGs) with CO5BOLD
\citep{Freytag2002AN....323..213F,
Chiavassa2009A&A...506.1351C,
Arroyo-Torres2015A&A...575A..50A}
show large-scale convection cells as well.
But for RSGs the ratio of surface pressure scale height, and therefore the size of the convective structures
and the local radius fluctuations, to radius is smaller than for AGB stars,
so that RSGs appear more spherical.

%===============================================================================
\subsection{Pulsations}\label{s:Pulsations}

%-------------------------------------------------------------------------------
\subsubsection{Exploring the pulsation mode and period}
\label{s:DerivePeriod}

%...............................................................................
\begin{figure*}
\centering
\includegraphics[width=18cm]{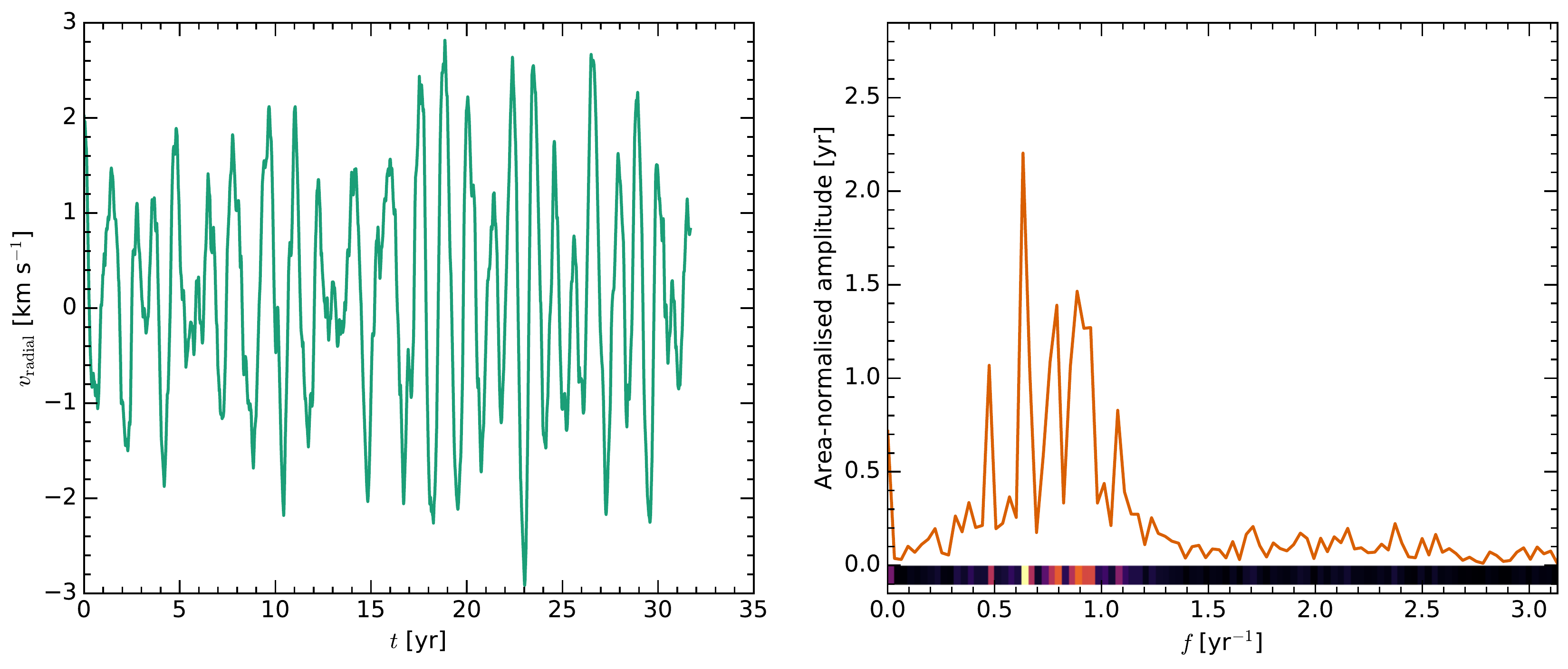}
\caption{Examples for the standard model st28gm06n26:
% Representation of the standard model st28gm06n26.
\textit{Left:} The vertical velocity at a constant $r$\,=\,301\,$R_\sun$, over 30 years.
\textit{Right:} The power spectrum of the vertical velocity,
shown to the left, showing the clearly dominant frequency.
At the bottom of the panel, the density plot for this power spectrum is shown.
This type of data is used in Fig.\,\ref{f:pow_comp_3mod} for all radial points.}
\label{f:vel_pow}
\end{figure*}
%...............................................................................

%...............................................................................
\begin{figure*}
\centering
\includegraphics[width=18cm]{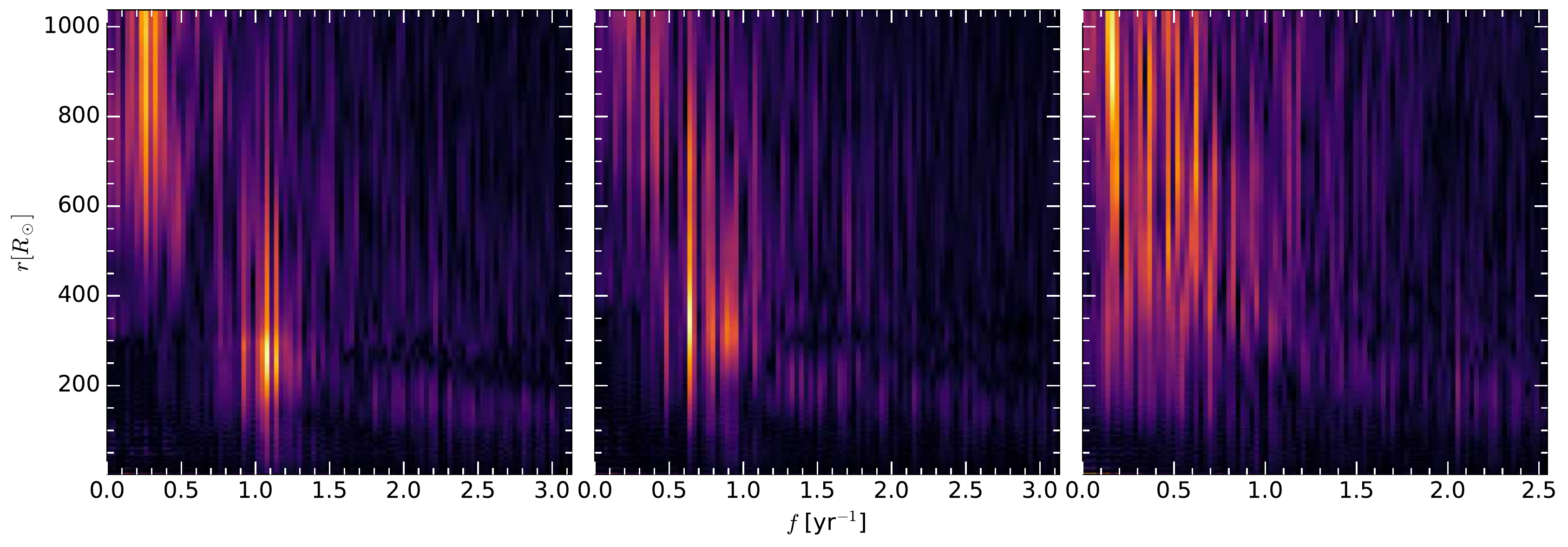}
\caption{Power spectra derived from the velocity fields of three different models,
mapped over frequency and radial distance.
For easier comparison all the different power spectra have been area normalized.
\textit{Left:} Model st29gm04n001, the hottest and smallest model, is shown.
\textit{Middle:} The standard model st28gm06n26 is shown.
\textit{Right:} Model st28gm07n001, the coolest and most luminous model with the largest radius, is shown.}
\label{f:pow_comp_3mod}
\end{figure*}
%...............................................................................

%...............................................................................
% --- aaagb2_intseq_plot.pro ---
\begin{figure*}[hbtp]
%\hspace*{0.9cm}\includegraphics[width=5.7cm]{aaagb2_st29gm04n001_rho_mean.pdf}
%\hspace{0.1cm}\includegraphics[width=5.7cm]{aaagb2_st28gm06n26_rho_mean.pdf}
%\hspace{0.1cm}\includegraphics[width=5.7cm]{aaagb2_st28gm07n001_rho_mean.pdf}\vspace{0.6cm}
\includegraphics[width=6.557cm]{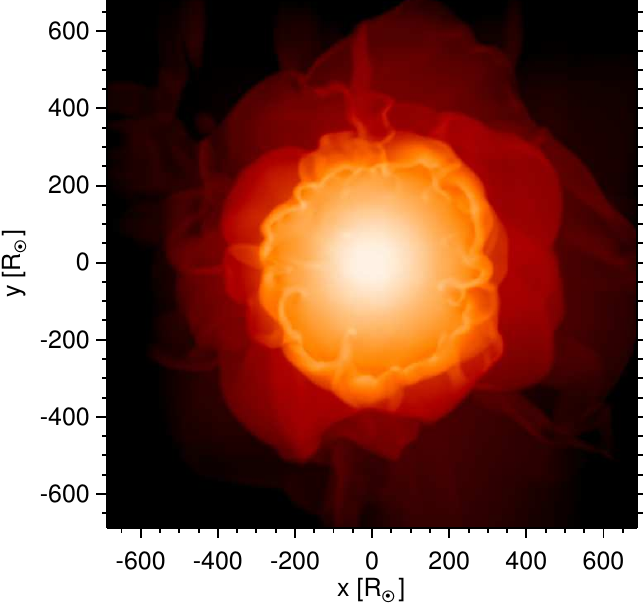}
\includegraphics[width=5.6cm]{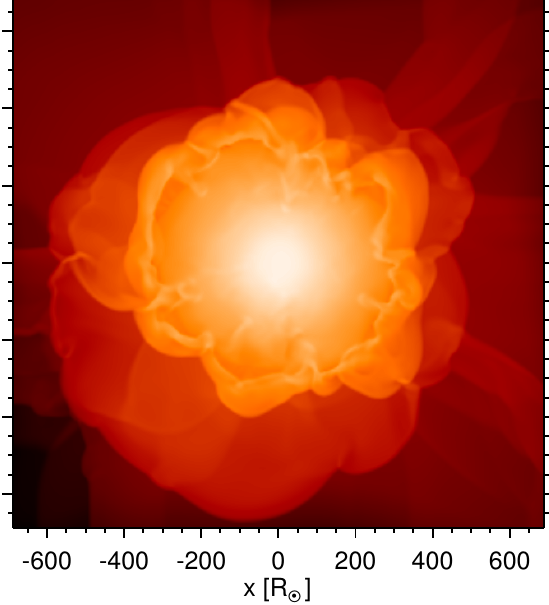}
\includegraphics[width=5.6cm]{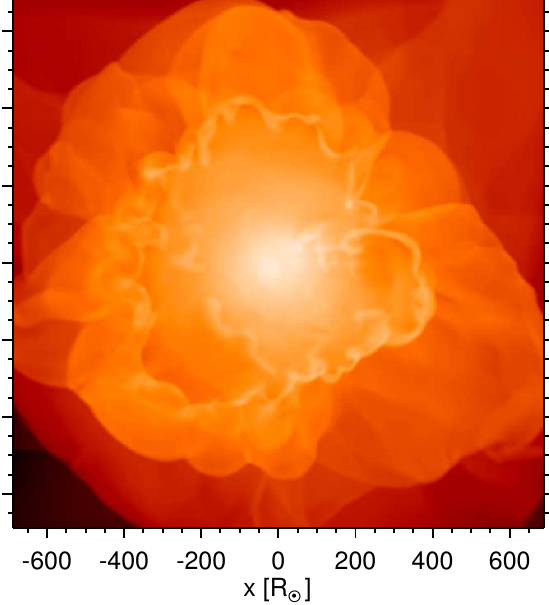}
\caption{Density slices for the three models used in Fig.\,\ref{f:pow_comp_3mod}.
The range used for all color tables is -16\,$\le$\,$\log \rho$\,$\le$\,-6.7.
}
\label{f:rho_comp}
\end{figure*}
%...............................................................................

The density snapshots in Fig.\,\ref{f:st28gm06n25_QuSeq} show irregular structures
with convection cells in the interior and a network of shocks in the atmosphere.
To investigate purely radial motions we consider the
density-weighted spherical average of the radial velocity
$\langle v_\mathrm{radial} \rangle(r,t)$
as described in Sect.\,\ref{s:Setup}
and plotted in Fig.\,\ref{f:vel_field} for the standard model st28gm06n26.

The behavior of the inner part of the model differs
from that of the atmosphere, which is particularly evident in Fig.\,\ref{f:vel_field}. 
Below about 400\,$R_{\sun}$
(the nominal radius is $R_\star$\,=\,$371\,R_\sun$),
the pulsation is rather regular and coherent over all layers,
close to a standing wave.
The fundamental mode dominates
as there are no nodes in the velocity map.
In the outer layers, however,
the slopes in the velocity map indicate propagating shock waves
(see Sect.\,\ref{s:Shocks}).

In the right panel in Fig.\,\ref{f:vel_field}, a close-up of part of
the velocity field is shown, with movements of mass shells overlaid,
analogous to plots for the 1D models, for example, in
\citet{Hoefner2003A&A...399..589H} or \citet{Nowotny2010A&A...514A..35N}.
The amplitude of the 3D mass-shell oscillations is smaller than
for corresponding 1D models by a factor of two or more.
This is at least partly due to the averaging over the full 3D model,
which smoothes the amplitude
as the shock waves are not exactly spherical.

To quantify the periodic behavior a Fourier analysis is performed
using the averaged radial velocity $\langle v_\mathrm{radial} \rangle(r,t)$.
An example for $r$\,=\,$301\,R_\sun$ is shown
in the left panel of Fig.\,\ref{f:vel_pow}, again for model st28gm06n26
with the corresponding power spectrum in the right panel.
Frequencies around 0.6\,yr$^{-1}$ dominate.

To explore the behavior all through the star
the power spectra of the averaged radial velocity $\langle v_\mathrm{radial} \rangle(r,t)$
at all radii are plotted
in the middle panel of Fig.\,\ref{f:pow_comp_3mod}.
Again the two different behaviors in the inner and outer layers of the model are evident.
There is a dominant frequency in the stellar interior,
as suspected from Fig.\,\ref{f:vel_field}.
However, the outer layers beyond $r$\,$\sim$\,$400\,R_\sun$ no longer
pulsate with the same period as the interior of the star.
Lower frequency signals become more prominent and beyond $r$\,$\sim$\,$600\,R_\sun$
(far out in the shock-dominated atmosphere), and
the frequency with the largest amplitude is significantly smaller
than that of the interior of the star.

In the power spectra of models with different stellar parameters
in Fig.\,\ref{f:pow_comp_3mod},
the dominant mode generally becomes more diffuse with increasing radius of the model.
For more compact models,
for instance model st29gm04n001 with $R_\star$\,=\,$294\,R_\sun$
in the left panel in Fig.\,\ref{f:pow_comp_3mod},
there is a very clear dominant mode in the interior.
For the standard model
(middle panel in Fig.\,\ref{f:pow_comp_3mod} with radius $R_\star$\,=\,$371\,R_\sun$)
there is still a dominant frequency, but the spread around this frequency is larger.
For the largest model
(model st28gm07n001 with $R_\star$\,=\,$531\,R_\sun$
in the right panel in Fig.\,\ref{f:pow_comp_3mod}),
the power spectrum seems to be equally distributed over a large frequency range,
lacking the clear dominant frequency that is present in the two other models.

To find the dominant frequency and therefore the pulsation period and to investigate the spread in frequencies,
the area-normalized power spectra of the radial velocities
for radial distances $r$\,=\,0.5\,--\,1\,$R_\star$ were added.
For instance, this corresponds to radial points
in the range $r$\,=\,185.5\,--\,371\,$R_\sun$ for the standard model.
A Gaussian distribution was fitted in the frequency domain containing the strongest signal. 
The central value for the fit for each model is taken as the dominant frequency $f_\mathrm{dom}$,
while the spread in frequencies is represented by the standard deviation $\sigma_\mathrm{puls}$.
The resulting periods $P_\mathrm{puls}$\,=\,$1/f_\mathrm{dom}$
and spreads $\sigma_\mathrm{puls}$
are listed in Table\,\ref{t:ModelParam} and plotted in Fig.\,\ref{f:pr_plum_comp},
where the colors of the squares represent the temperature
with lighter colors indicating higher temperature
and the size represents the luminosity of the model, as seen in Fig.\,\ref{f:overview_models}.

%-------------------------------------------------------------------------------
\subsubsection{Excitation mechanism}

\begin{figure*}[htbp]
\hspace{-1.5mm}\includegraphics[width=9.1cm]{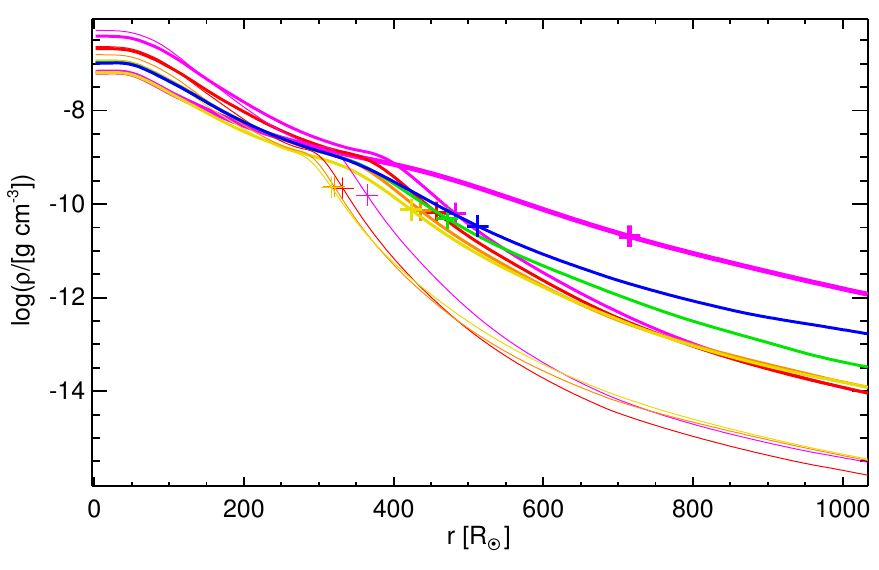}
\includegraphics[width=9.1cm]{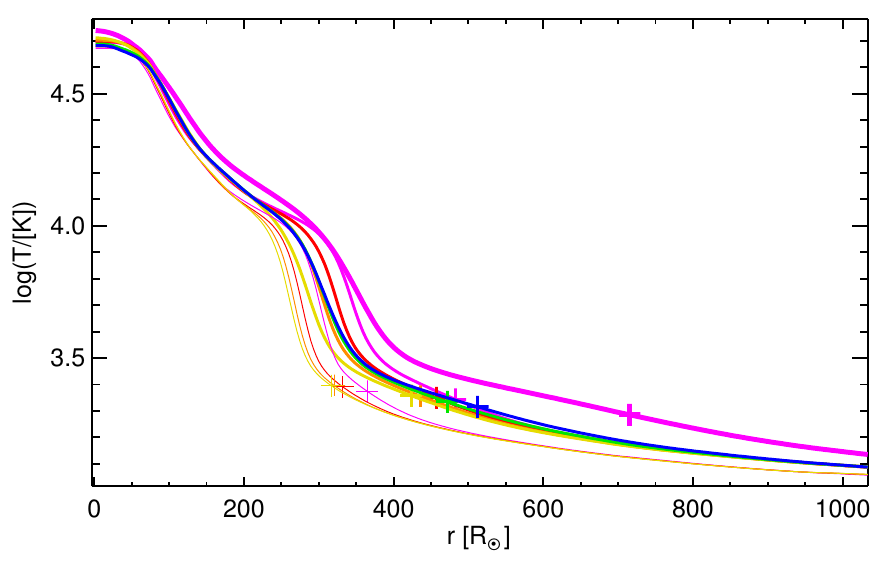}\hspace{-2.5mm}

\vspace{-2mm}
\hspace{-1.5mm}\includegraphics[width=9.1cm]{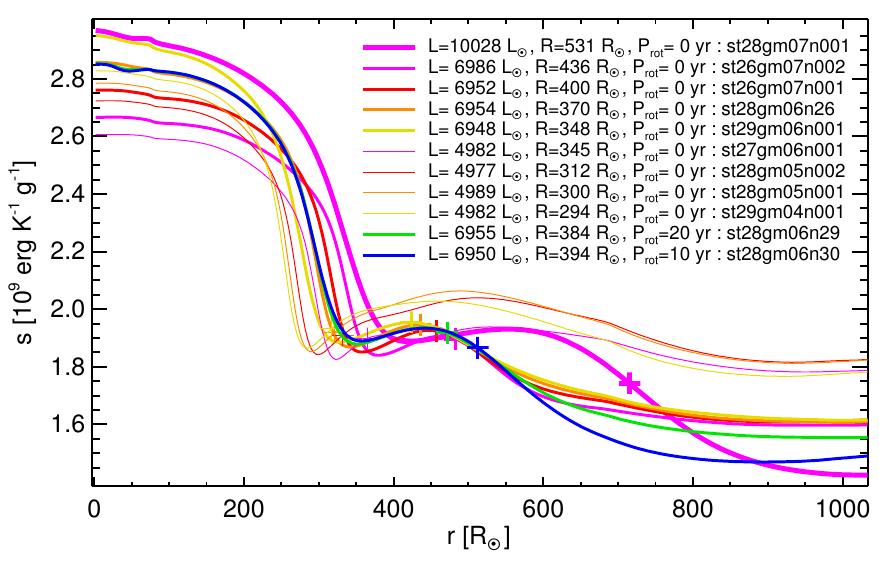}
\includegraphics[width=9.1cm]{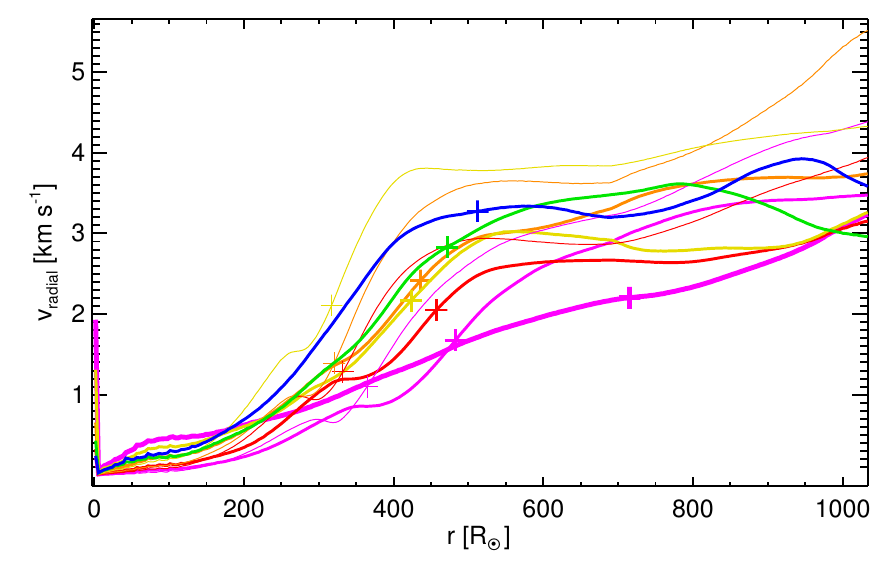}\hspace{-2.5mm}

\vspace{-2mm}
\hspace{-1.5mm}\includegraphics[width=9.1cm]{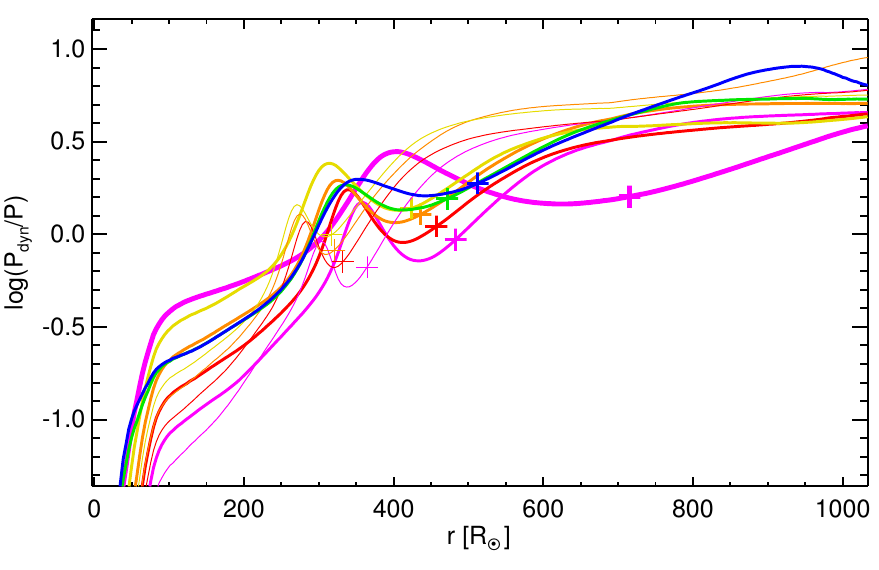}
\includegraphics[width=9.1cm]{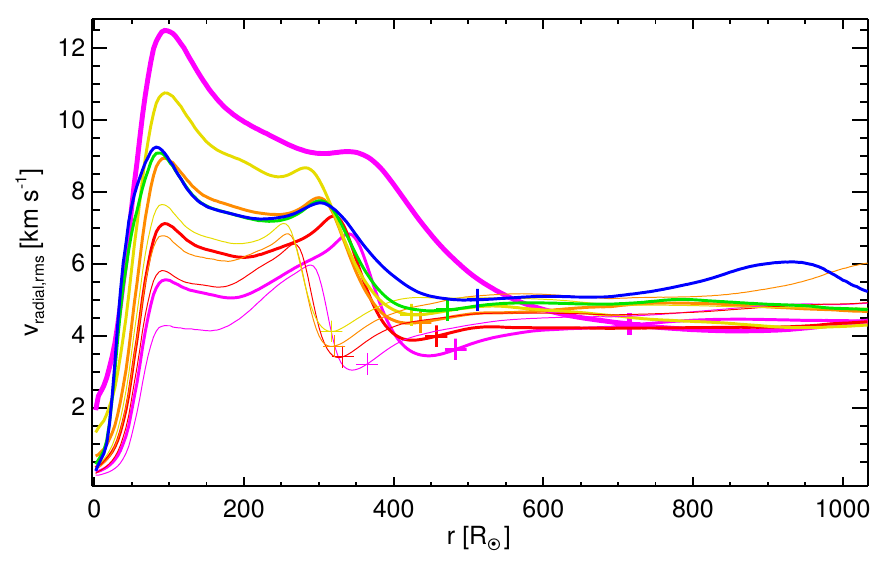}\hspace{-2.5mm}

\caption{Plots of selected quantities, averaged over spherical shells and time
vs. radii for all grid models.
The plus signs roughly indicate the depth with $\tau_\mathrm{Ross}=1$.
\textit{Top left:} Logarithm of the mass density~$\rho$.
\textit{Top right}: Logarithm of the temperature~$T$.
\textit{Center left}: Entropy per mass unit~$s$.
\textit{Center right}: Temporal rms value of the spherically averaged radial velocity
(includes mostly contributions from the radial pulsations).
\textit{Bottom left}: Logarithm of the ratio of dynamical pressure~$P_\mathrm{dyn}$ and gas pressure $P$.
\textit{Bottom right}: Temporal and spatial rms value of the radial velocity
(includes contributions both from convection and radial pulsations).}
\label{f:qoverR}
\end{figure*}

The irregular spread of the mode frequencies in Fig.\,\ref{f:pow_comp_3mod}
is likely due to interactions between the pulsations and large-scale convective motions
causing occasional amplitude changes and phase shifts.
With larger radii, the convective cells increase further in relative size
resulting in stronger disturbances of the pulsation mode. 
Within a luminosity group the frequency spread grows with decreasing radius,
which is likely due to the increase of convective velocities
with increasing effective temperature;
cf.\ the bottom right panel in Fig.\,\ref{f:qoverR}.

The power spectra in Fig.\,\ref{f:pow_comp_3mod} can be
compared to the power spectrum derived from a local solar model
in Fig.\,3 of \cite{Stein2001ApJ...546..585S}.
The similarity is an indication of a common excitation mechanism.

Analyzing light curves,
\cite{Christensen-Dalsgaard2001ApJ...562L.141C}
attributed oscillations in semi-regular variables to stochastic excitation by convection.
\cite{Bedding2005MNRAS.361.1375B} distinguished several cases.
These authors attributed large phase fluctuations in the light curve of the semi-regular star W~Cyg
to stochastic excitation,
whereas the very stable phase of the true Mira star X~Cam was found to be consistent with
the excitation by the $\kappa$ mechanism.
L$_2$~Pup was classified as an intermediate case in which both mechanisms play a role.

Analyzing the work integral in 1D pulsation models,
\cite{Lattanzio2004.ENPoAGB..23} concluded
that the driving mechanism, at least for Mira variables, is likely a $\kappa$~mechanism
acting in the partial hydrogen and helium~I ionization zone.

Our grid of 3D models, which only covers a small part of the entire range of the AGB star parameters,
already shows a range of different behaviors of the oscillations
(see Fig.\,\ref{f:pow_comp_3mod} and the discussion in the previous section),
where the trend clearly points to the role of convection and the size of
the convective cells for the excitation of, or at least interaction with, the pulsation.
The non-stationary transonic convective flows
with Mach numbers, which in the downdrafts often exceed unity,
produce acoustic noise
as described by \cite{Lighthill1952RSPSA.211..564L}
for turbulent flows.
This excitation mechanism has been investigated
for instance by \cite{Nordlund2001ApJ...546..576N} and \cite{Stein2001ApJ...546..585S}
with local 3D radiation-hydrodynamics simulations of solar granulation.
The Mach numbers in near-surface convective flows of AGB stars can be even larger
causing more efficient wave excitation
\citep{Lighthill1952RSPSA.211..564L}.
As the relative sizes of the exciting convective structures are very large,
the generated waves have much lower wave numbers than on the Sun.

To check that the radial pulsations are not just (long-lasting) transient
phenomena introduced by the initial conditions, we added a strong (purely
artificial) drag force in the entire model volume.
This drag force reduced the amplitude of the pulsations (and the convective flows)
but did not lead to an exponential decay of the radial mode,
indicating that an efficient mode-excitation mechanism must be at work.

In the models of \cite{Freytag2008A&A...483..571F},
these pulsations existed and their amplitude was extracted as a description of
the piston boundary in 1D models.
However, the global pulsations were harder to distinguish from the local shock network
because of the lower numerical resolution;
this led to larger sizes of convective and wave structures
and shorter time sequences, which made a Fourier analysis less reliable
than what is possible for the current model grid.

%-------------------------------------------------------------------------------
\subsubsection{Comparison with 1D models and observations}

%...............................................................................
\begin{figure*}
\centering
\includegraphics[width=18cm]{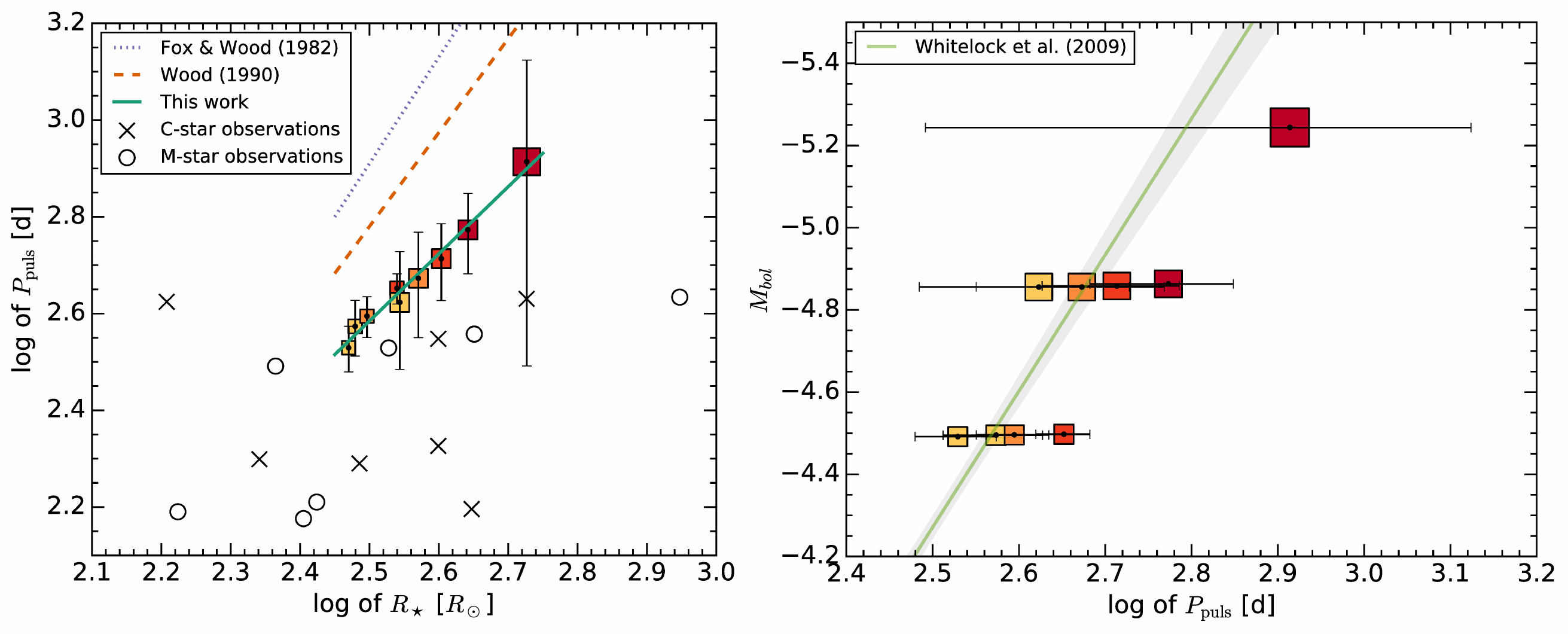}
\caption{\textit{Left:} Logarithm of the period of the models plotted
against the logarithm of the resulting radius
with three different period-radius relations plotted.
The green line indicates period-radius relations from this work,
the orange dashed line indicates  \cite{Wood1990fmpn.coll...67W},
and the purple dotted line indicates \cite{Fox1982ApJ...259..198F} for $M$\,=\,1\,$M_\sun$.
The error bars of the pulsation period of the models refer
to the width of the peak in the power spectrum,
that is, larger than the uncertainty of the period.
The statistical error in the model radius is small.
The typical error in the observed radii is about 31\,\%
due to uncertainties in the parallax.
The crosses and circles are observations of C~stars and M~stars, respectively,
with radius from the CHARM2 catalog \citep{Richichi2005A&A...431..773R} 
and periods from GCVS catalog \citep{Samus2009yCat....102025S}.
Only stars with measured parallaxes were picked
so that the  distance determination is independent of the measured period \citep{Ramstedt2014A&A...566A.145R}. 
\textit{Right}: The absolute magnitude against the logarithm of the
period for all models.
The error in the absolute magnitude is only due to the finite simulation time
and is tiny (a few mmag).
The line is taken from \cite{Whitelock2009MNRAS.394..795W}
and is the P-L relation for AGB stars in the LMC, 
where the gray area is the 1~$\sigma$ error of the fit to the observations.
% \LEt{Please see Sect. 1.3.2 in Guide.
% Please avoid discursive
% language in the caption as well as repeated
% informational text in the body of the paper. Please check for this throughout
% as I will not note these.}
}
\label{f:pr_plum_comp}
\end{figure*}
%...............................................................................

As has been pointed out by
\cite{Fox1982ApJ...259..198F}
and
\cite{Wood1990fmpn.coll...67W},
AGB stars do not seem to follow
the simple period-mean-density relation, $P_\mathrm{puls}\times(\bar{\rho} / \bar{\rho}_\odot)^{1/2} = Q$,
where $Q$ is the pulsation constant.
This is not very surprising
because the derivation of the period-mean-density relation
relies on the assumptions
that displacements are adiabatic and non-linear effects are small,
both of which are probably incorrect for AGB stars. 
Instead, these works, using 1D pulsation models,
find that $P_\mathrm{puls}$\,$\propto$\,$R_\star^\alpha M_\star^{-\beta}$
with $\alpha$\,$\sim$\,1.5\,--\,2 and $\beta$\,$\sim$\,0.5\,--\,1. 
The period-radius relation
for our models with a fixed mass $M_\star$\,=\,1\,$M_\sun$
is compared to that of
\cite{Fox1982ApJ...259..198F}
and
\cite{Wood1990fmpn.coll...67W}
in the left panel of Fig.\,\ref{f:pr_plum_comp}.
We find
\begin{equation}
  \log(P_\mathrm{puls}) = 1.39 \, \log(R_\star) - 0.9
  \enspace ,
\end{equation}
which gives generally a larger radius for a given period.
There might be several reasons for this difference
in addition to uncertainties in the 1D models.

There is a contribution to the extension of the atmosphere of the 3D models
(see Sect.\,\ref{s:DynamicalPressureAndRadius})
due to the convectively generated, small-scale shocks (see Sect.\,\ref{s:Shocks})
that do not exist in the 1D models.
This would affect the photospheric radius but not the pulsation period
if the shocks occur above the top of the acoustic cavity.
This contribution
% \LEt{Please specify what "it" refers to if this is not correct.}
might even lower the slope
because the largest models have the most extended atmospheres.
The convective envelope
is not in hydrostatic equilibrium
but is affected by the convective dynamical pressure as well.
In addition, the treatment of the artificial core in the 3D models might play a role.
However, the effect should be small because the sound-crossing
time for the core, and therefore the contribution to the period, is relatively small.
Finally, there is some uncertainty in the determination of radius and period.

In the left panel of Fig.\,\ref{f:pr_plum_comp},
observations of the radii for different AGB stars
are plotted against the periods
with C~stars shown as crosses and M~stars as circles. 
The diameter observations are from \cite{Richichi2005A&A...431..773R},
periods from \cite{Samus2009yCat....102025S},
and parallaxes from \cite{vanLeeuwen2007A&A...474..653V}.
The 3D models agree fairly well with observations,
especially the lower luminosity models. 
The period-radius relation from the 1D models might produce periods that are too long for the
range of radii explored by the 3D models when compared to observations.
However, 1D models predict that altering the mass
leads to a change of the period-radius relation.
While only the relations for 1\,$M_\sun$ are plotted here
for direct comparison to the results from the 3D models,
1D models from \cite{Fox1982ApJ...259..198F} can reach the shorter periods,
if the mass of the models is increased.

It is difficult to draw any final conclusion as the
errors in determining fundamental parameters for field AGB stars are
very large. The uncertainties in observed absolute magnitudes
originate mainly from uncertainties in the parallaxes, which
are difficult to determine for AGB stars as the photo-centers of
these stars are variable
\citep[for a statistical analysis of photocentric variability, see][]{Ludwig2006A&A...445..661L,Chiavassa2011A&A...528A.120C}.
Also, the uncertainty of
% radii observation for
observations of the radii of
AGB stars is fairly large, sometimes on the order of the
stellar radius. In addition, the radius varies significantly during a pulsation
cycle and the phases are not always well known.
Here, we use the mean radius as reference.
% Here, we compare to the mean radius.
% \LEt{Please specify what you are comparing to the mean radius, for example, "the modeled radius...".}
However, the radius varies by
around 20\,\% during one pulsation period for our models.

It is also likely that the stellar masses affect the
period-radius relation. However, unless there are
well-observed companion stars, it is usually not possible to
determine masses from observations. A theoretical prediction
of the effect of different masses is not yet possible, as
all the models in the current grid have a mass of 1\,$M_\sun$.

A property that is better constrained by observation is the
P-L relation, which has been extensively studied.
A comparison between the 3D models and observation by \cite{Whitelock2009MNRAS.394..795W}
is shown in the right panel of Fig.\,\ref{f:pr_plum_comp}. The models from the
grid follow a trend of brighter absolute magnitude with larger period,
which is qualitatively similar to that of the observations.
There is, however, a spread in the periods because of the different effective temperatures and radii for a given luminosity,
giving constraints for our -- so far a bit arbitrary -- choice of the
combination of the main control parameters $M_\star$, $M_\mathrm{env}$, $L_\star$, and metallicity
of the 3D models.

%...............................................................................
% --- aaagb2_seqavgstat_plot.pro ---
%...............................................................................

%===============================================================================
\subsection{Atmospheres with networks of shocks}\label{s:Shocks}

The steepening of acoustic waves
in the solar chromosphere
and the transformation into a network
of shocks was modeled by
\cite{Wedemeyer2004A&A...414.1121W}
and \cite{Muthsam2007MNRAS.380.1335M}.
AGB stars have a larger temperature drop from the convection zone to the atmosphere
\citep[cf.\ Figs.\,5 and 6 in][for a model sequence
from the Sun with $\log g$\,=\,4.44
to a 5\,$M_\sun$ star with $\log g$\,=-0.43]{Freytag2013EAS....60..137F} that is
accompanied by a larger change in pressure or density scale height.
This leads to a stronger compression and amplification of the waves in the cool atmosphere,
so that the waves very early turn into shocks,
leaving no room for an essentially undisturbed photosphere.
Sound-speed variations,
particularly at the rugged surface of the convection zone,
and transonic convective flow speeds
shape the waves and contribute to the generation of small-scale shock structures
(at $r$\,$>$\,300\,$R_\sun$ in Fig.\,\ref{f:vel_field_1}),
which give rise to ballistic gas motions
with peak heights of a few ten to a few hundred solar radii.
The strongest shocks can even leave the computational box.

The plot of entropy versus\ radius in Fig.\,\ref{f:qoverR} shows
that part of the atmosphere is a zone of convective instability --
with negative entropy gradient --
separate from the normal interior convection zone.
The image sequences in Fig.\,\ref{f:st28gm06n25_QuSeq}
and the plot of $v_\mathrm{radial}(r,t)$ in Fig.\,\ref{f:vel_field_1}
show a very different behavior of the two zones:
only the inner zone is governed by the normal overturning motions,
whereas propagating shocks dominate in the outer zone
(as discussed in Sect.\,\ref{s:DerivePeriod}).
But still, the convective instability might destabilize the shocks
favoring a tendency toward smaller structures as seen in
the intensity snapshots
in Figs.\,\ref{f:st28gm06n25_QuSeq} and \ref{f:st28gm06n26_IntSeq}.
For models with $L$\,=\,5000\,$L_\sun$, this instability zone lies two times further out
relative to the stellar radius, compared to the higher luminosity models.
\cite{Woitke2006A&A...452..537W} showed that in 2D simulations of the atmosphere
of an AGB star,
where the radial stellar pulsation and 
inhomogeneities generated by convection are ignored
and shocks are generated by an external $\kappa$~mechanism,
instabilities within the shock fronts
can cause small-scale structures to form.

A combination of relatively small-scale acoustic noise and global, radial pulsations
generates the network of shocks
with a size spectrum regulated by several complex processes;
a wave emitted from a small region close to the surface
might turn into an expanding shock wave,
which fills a large part of the atmosphere if the surrounding flow permits it.
The overtaking and merging of shocks increases the typical size.
On the other hand, large-scale waves in the interior are shaped by
the convective background flow and sound-speed distribution and can become
as rugged as the surface of the convection zone itself.
And the convective instability in the atmosphere also favors small scales.

The shock fronts show up prominently in the density slices in Fig.\,\ref{f:st28gm06n25_QuSeq}
and are only occasionally visible in the temperature slices
(e.g., in the upper left quadrant in the fourth row)
because the radiative relaxation is fast enough to cool down the heated post-shock material to the local equilibrium value, even with the gray radiative transfer used.
The radiative relaxation time decreases further when
non-gray radiative transfer is employed
(in a first simulation),
essentially wiping out any temperature signature of the shocks.
That might change again with a drastic increase in numerical resolution,
which is currently not achievable.

While the temperature stratifications of the various models in Fig.\,\ref{f:qoverR}
show many differences in the convection zone and inner atmosphere, 
they converge into one of three profiles in the outer atmosphere,
depending only on stellar luminosity.
This and the relatively small temperature fluctuations in the outer part
of the temperature slices in Fig.\,\ref{f:st28gm06n25_QuSeq}
indicate
that the thermal structure of the outer atmosphere
is dominated by radiative processes and is not far from radiative equilibrium.
The shocks show up in the velocities and the density but hardly in the temperature.

Expectedly, the size spectrum of the shocks is extended
toward smaller scales with increasing resolution
as a comparison of Fig.\,\ref{f:st28gm06n25_QuSeq} of the current paper
with Fig.\,1 of \cite{Freytag2008A&A...483..571F} demonstrates.

%===============================================================================
\subsection{Rotation}

%...............................................................................
% --- aaagb2_rotslice_plot.pro ---
\begin{figure}[hbtp]
\includegraphics[width=4.45cm]{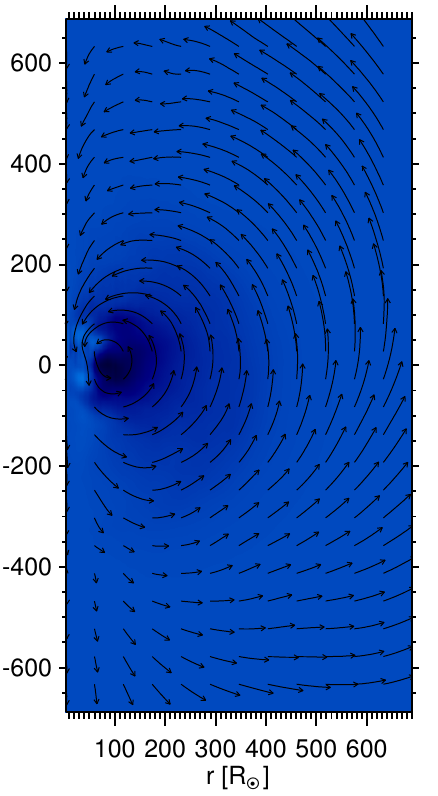}
\includegraphics[width=4.45cm]{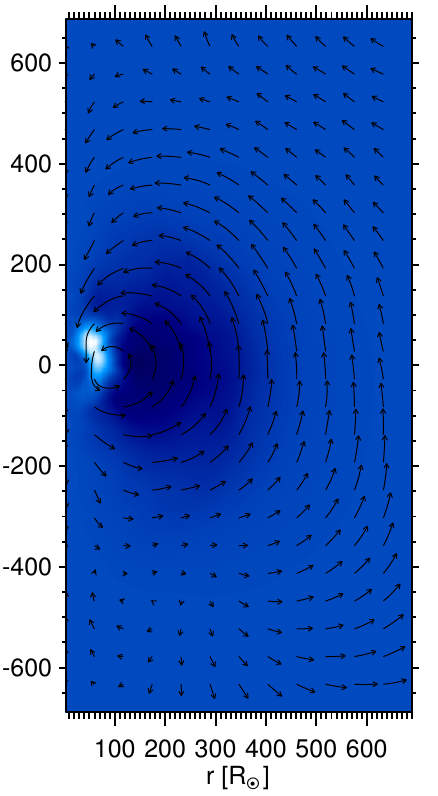}\hspace*{-2mm}
\caption{Azimuthally and temporally averaged velocity fields for models
st28gm06n29 ($P_\mathrm{rot}$\,=\,20\,yr, left)
and
st28gm06n30 ($P_\mathrm{rot}$\,=\,10\,yr, right).
The color plot shows the angular momentum in the corotating frame;
bright blue means rotation faster than the mean given by $P_\mathrm{rot}$,
and dark blue means rotation slower than the mean.
}
  \label{f:st28gm06n29_rotslice}
\end{figure}
%...............................................................................

Rotation plays an important role in stellar activity and possibly
also in the dynamics of stellar winds.
The loss of angular momentum
from magnetic coupling with the surroundings
or from a wind
likely slows down evolved stars considerably.
\cite{Dorfi1996A&A...313..605D} performed 1D stationary-wind models
of a 1\,$M_\sun$/2600\,K/10000\,$L_\sun$ AGB star
and found that a rotation period of 40\,yr modifies the isotropic mass loss marginally,
while a period of 10\,yr results in a drastic increase of the mass loss rate
and causes a significant axial asymmetry of the wind.

% Hydrodynamical simulations of convection in rotating stars of the solar interior
% or stellar interiors  have been performed,
%% Hydrodynamical simulations of convection in the interior of rotating stars have been performed.
%% In these simulations, the Mach numbers are low, so that the flow is essentially incompressible
In hydrodynamical simulations of convection in the interior of rotating stars,
the Mach numbers are so low that the flow is essentially incompressible
and the optical thickness of individual grid cells is so large
that radiation transport is adequately described by the diffusion approximation.

However, these conditions are not met
in the convective envelopes and surrounding atmospheres of AGB stars,
where the numerical treatment in CO5BOLD
(e.g., of non-local radiation transport and compressible hydrodynamics)
is required,
so we modified the CO5BOLD setup to account for rotation.
While it is possible to
just add
% add just
% simply add
% add
a rotational velocity field to the initial model,
we chose to perform the simulation in a corotating frame.
A centrifugal potential is added to the gravitational potential.
Before each hydrodynamics step the Coriolis force is applied,
which just rotates the velocity vectors, but with twice the amount simply
suggested by period and time step.
The artificial drag force in the model core was chosen to act only radially
so as not to affect the angular momentum.
We computed a first exploratory pair of models with rotation periods of 20 and 10 years
(see Table\,\ref{t:ModelParam}).
Longer periods would require even longer integration times.
The $P_\mathrm{rot}$\,=\,20\,yr simulation was started from a snapshot
from a non-rotating run,
% and the periods (i.e., centrifugal potential and Coriolis force) were increased
and the period was decreased (by increasing centrifugal potential and Coriolis force)
gradually over a few stellar years to avoid transient oscillations due to a too rapid
change of the potential.

As expected for slow rotators,
where the rotational period
is longer than typical convective turnover timescales,
angular momentum is advected inward
into a small region close to the core of the model
(see Fig.\,\ref{f:st28gm06n29_rotslice}).
In spite of the drag force in the core,
a global dipole flow develops
with typical velocities in the atmosphere of
$v_\mathrm{meridional}$\,=\,4\,km/s (for st28gm06n29 with $P_\mathrm{rot}$\,=\,20\,yr)
and
$v_\mathrm{meridional}$\,=\,2.5\,km/s (for st28gm06n30 with $P_\mathrm{rot}$\,=\,10\,yr).

While the core generally rotates very rapidly,
the part of the convection zone that is not close to the axis
rotates slower than the nominal rate.
Part of the material close to the core of the model with  $P_\mathrm{rot}$\,=\,20\,yr
even shows a slow retrograde rotation.
However, all azimuthally averaged velocity components show large fluctuations
and the averaging time or relaxation time for this model might not yet be sufficient.

The mean atmospheric stratification is affected by the rotating star:
while the temperature stratification shows hardly any effect,
the average density in the atmosphere increases with shorter rotation period
(see Fig.\,\ref{f:qoverR}).

Our rotating models have a number of shortcomings.
The approximation of the smoothed stellar core plays a larger role than
for purely convective (and pulsating) flows that are not rotating.
Would the angular momentum in a real star be advected even further in
and leave only a very slowly rotating convective envelope and atmosphere behind?
What role would magnetic fields play in coupling the interior to the
convective envelope?
The outer boundaries might also influence the results because the slowly rotating atmosphere moves with respect to the
computational box and might exchange angular momentum through the boundaries.
Clearly, improved models need a larger computational domain and
update of the treatment of the stellar core.

Still, the presented models support the results of
\cite{Dorfi1996A&A...313..605D}
that rotation in AGB stars -- if the period is on the order of 20\,yr or shorter --
can influence the atmospheric velocity field and the wind,
and might be responsible for the shapes of some planetary nebulae.

%#########################################################################################
\section{Discussion}

%===============================================================================
\subsection{Influence of numerical parameters}

%...............................................................................
% --- aaagb2_seqavgstat_plot.pro ---
\begin{figure}[htbp]
\includegraphics[width=8.8cm]{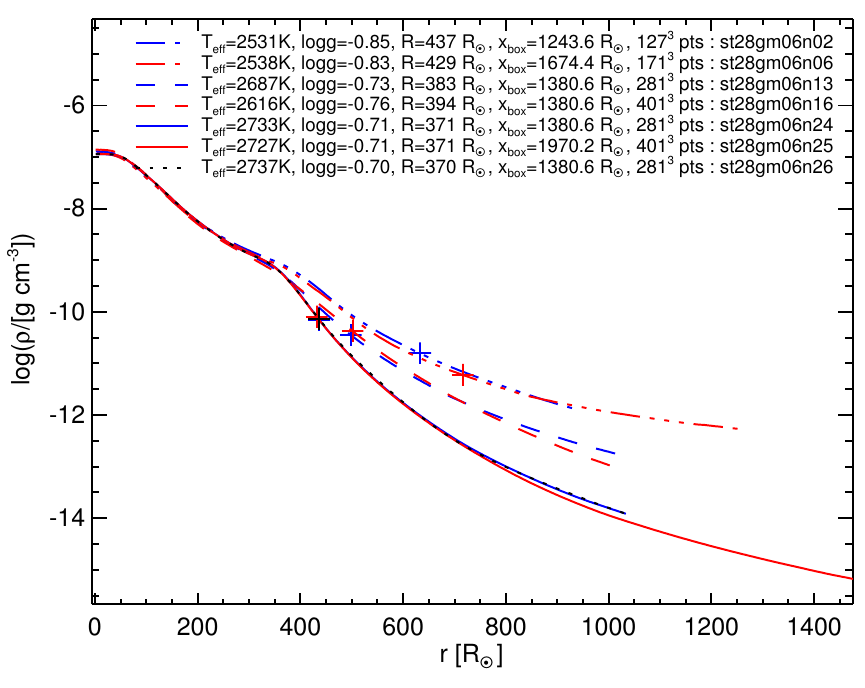}
\caption{
Logarithm of the gas density averaged over spherical shells and time
for six models.
The ordering in the legend essentially corresponds to the order of the curves.
The plus signs roughly indicate the depth with $\tau_\mathrm{Ross}=1$.
In the legend,
the effective temperature,
surface gravity,
edge length of the computational box,
number of grid points,
and the name
are given for each model.
}
\label{f:st28gm06nXX_rhoOverR}
\end{figure}
%...............................................................................

The current models have an improved resolution over those presented in
\cite{Freytag2008A&A...483..571F}
because of an increased number of grid points and a higher order reconstruction scheme
of the hydrodynamics solver \citep{Freytag2013MSAIS..24...26F}.
While sub-surface convection cells and above-surface shocks show finer structures,
there are no qualitative changes in the general results.
Still, the newer models allow a better separation of the radial pulsation mode
from the convective noise
and also present a longer time base for the Fourier analysis.

The averaged density stratifications for selected models
in Fig.\,\ref{f:st28gm06nXX_rhoOverR}
give an idea about the size of effects due to changes in numerical parameters.
The two smallest and oldest models, n02 and n06
(referring to the last three letters of the model names),
 which were already used in \cite{Freytag2008A&A...483..571F},
have the same resolution but different extensions of the computational box
and agree well with each other.
The same is true for the pair n24/n25 from the current ``test'' group of models,
indicating that the box size does not have a major impact.
The curves for
% n24 and n26 (our standard model)
n24 and our standard model n26
are indistinguishable.
The difference is a change in the version of the  code,
which mainly entails
a modification of the WENO scheme used for the high-order reconstruction
in the hydrodynamics solver.
We implemented a reduction of the order of the reconstruction,
and therefore an increase in dissipation, in case of large Mach numbers
or large gradients in pressure or entropy.
The numerical resolution of model n16 was increased by a factor of
1.43 in each direction, compared to n13 with only a slight effect on the mean-density stratification.

The noticeable decrease of the density by about one order
of magnitude in the outer layers from n13/n16
to n24/n25/n26 is caused by a change in the outer boundary conditions.
Assuming a less steep density decrease
while filling the ghost cells at the boundaries of the computational box
induces a stronger infall of material in the phase between outward
moving shocks and somewhat impedes propagation of the shocks, leading to lower
averaged densities in the outer layers.
An additional reduction might be due to a
difference in the treatment of the velocity damping in the core region of the models,
which prevents large-scale dipolar flows that
could develop and dominate the entire convective envelope.
Between the old models n02/n06 and the new models,
there are differences in the envelope mass $M_\mathrm{env}$
and too many changes in the numerics
to allow a disentanglement of the influence of individual settings.

%===============================================================================
\subsection{Dynamical pressure}\label{s:DynamicalPressureAndRadius}

The bottom left panel in Fig.\,\ref{f:qoverR}
shows the ratio of
the radial component of the dynamical pressure and the gas pressure,
both averaged over spheres and time,
$\langle \rho v_\mathrm{radial}^2 \rangle_{\Omega,t} / \langle P \rangle_{\Omega,t}$.
From the total dynamical pressure, we derive a density-weighted rms value of the radial velocities
according to
$(\langle \rho v_\mathrm{radial}^2 \rangle_{\Omega,t}/\langle \rho \rangle_{\Omega,t})^{1/2}$,
and the (smaller) contribution by the radial pulsations
$(\langle (\langle \rho v_\mathrm{radial} \rangle_{\Omega}^2 / \langle \rho \rangle_{\Omega}) \rangle_t / \langle \rho \rangle_{\Omega,t})^{1/2}$,
plotted in the bottom right and the middle right panels, respectively.

In the convection zone, where the pressure ratio is below one,
% it already stays above 20\,\% for a number of models for quite a fraction of the radius.
it already stays above 20\,\% for a number of models for a large fraction of the radius.
Thus, the dynamical pressure is not negligibly small and
influences the stratification at least of the outer convective envelope.
The peak of the radial velocities near the surface of the convection zone
is accompanied by a peak in the pressure ratio
and the dynamical pressure becomes even larger than the gas pressure.

Further out in the atmosphere, the dynamical pressure dominates over the gas pressure
(radiation pressure is not included in the current models)
by factor of 5 to 10.
It significantly increases the density and pressure scale heights
compared to the case with only gas pressure,
causing a levitation of dense material into cool layers.
This allows molecules to form
and possibly creates a highly irregular, non-spherical, and dynamic version
of a MOLsphere
\citep[an extended sphere around the star that is optically thick in molecular lines,
for instance, due to water; see][]{Tsuji2000ApJ...540L..99T}.
In addition, the conditions become even sufficiently cool and dense for dust to form
\citep[see, e.g.,][]{Freytag2008A&A...483..571F}.
The contributions of purely radially symmetric motions (middle right panel in Fig.\,\ref{f:qoverR})
and spatially fluctuating flows
to the total radial velocities (bottom right panel in Fig.\,\ref{f:qoverR})
are both significant, but 
only the radially symmetric motions can be accounted for by dynamical 1D models.

While we get very extended atmospheres in our models of AGB stars,
\cite{Arroyo-Torres2015A&A...575A..50A}
concluded that for red supergiant stars
(with much larger masses in the range 5\,--\,25\,\,$M_\sun$)
the dynamical pressure in
pulsating 1D models and the even larger dynamical pressure in
3D CO5BOLD models is not sufficient to enlarge the
photosphere to the observed sizes.

%===============================================================================
\subsection{Small-scale structures in the atmosphere}\label{s:FineStructures}

We presented some observations (e.g., by VLT/SPHERE, HST,
VLTI, and various other interferometers,
or with the Cassini spacecraft)
of asymmetries and clumps
in the dust envelopes of near-by AGB stars in the Introduction.

The bolometric-intensity maps in Figs.\,\ref{f:st28gm06n25_QuSeq} and \ref{f:st28gm06n26_IntSeq}
derived from the 3D~models
show that the smallest scale patterns change on timescales of less than a month,
while intensity changes of larger areas occur on timescales of about a year.

The surface of the normal stellar convection zone
(see Sect.\,\ref{s:Convection})
sits too deep to directly affect the emergent intensity:
% we do not see surface granules or larger convection cells.
surface granules and larger convection cells themselves are not observable.
Instead, the visible structures are caused by shocks on various scales.
However, as described in Sect.\,\ref{s:Shocks}, the shocks are shaped
by the underlying convective structures.
A dimming and brightening of a large area (see Fig.\,\ref{f:st28gm06n26_IntSeq})
might well indirectly reflect the dynamics of the convection.

A detailed comparison of results from the 3D models with observations
has to await simulations with non-gray opacities and a detailed treatment of dust
as well as time series of high-angular-resolution observations.

%===============================================================================
\subsection{Characteristic timescales}\label{s:TimeScales}

A comparison of the pulsation period with the various typical timescales
of convective structures (see Sect.\,\ref{s:Convection}) gives
\begin{equation}
  t_\mathrm{granule} \ll t_\mathrm{downdraft}
                   \sim \frac{\scriptstyle 1}{\scriptstyle 2} \, t_\mathrm{turnover}
                   \lesssim P_\mathrm{puls}
                   \ll t_\mathrm{giant-cell} 
  \enspace ,
\end{equation}
i.e., that the granules adjust to the pulsation
whereas the giant cells are more or less frozen in.
However, a restructuring of the giant cells can lead to a variation
of the detailed pulsation behavior on rather long timescales of several years.
The strongest interaction is to be expected between pulsations and downdrafts.
There is no such thing as ``the'' convective timescale.

%#########################################################################################
\section{Conclusions}

We presented a first exploratory grid of 3D radiation-hydrodynamics models of
AGB stars computed with a new version of CO5BOLD, which is improved in terms of
accuracy, stability, and boundary treatment compared to the version used for
the simulations presented in \cite{Freytag2008A&A...483..571F}.
The increased effective resolution leads to additional finer structures in the convective flow
(surface granules and deep turbulent eddies)
and in near-surface shocks.
However, there is no significant change in the dynamical behavior of the models
and only a small one in
quantities that are averaged spatially (over spheres) and temporally.

Several interacting processes govern the dynamics of the atmosphere
of the model AGB stars:
non-stationary convection manifests
as giant convective cells, which change topology on very long timescales,
and short-lived small surface granules.
The giant convections cells are outlined by turbulent downdrafts
that reach from the surface of the convection zone down into the very core of the model.
Convective motions emit acoustic noise
(i.e., waves, that get trapped inside the star, giving rise to standing acoustic modes,
comparable to solar p~modes)
and they shape, i.e., distort, wave fronts.
In addition, there are
large-amplitude, radial, fundamental-mode pulsations.
The small-scale acoustic waves steepen
when they reach the thin cool atmosphere and turn into shocks.
The shocks interact and merge,
so that the scale of the atmospheric shocks increases with radial distance,
from a small-scale shock network close to the surface of the convection zone
to distorted but almost global, more or less radially expanding shock fronts
in the outer layers.
The cycles of outward moving shocks and material falling back toward the star
have a longer period than the pulsations themselves.
% \LEt{As mentioned earlier, please do not use italics for emphasis.
% Check for these throughout and remove.}

The radial pulsations have realistic properties in spite of the
crude treatment of the stellar core
in the models.
The models reproduce the correct period for a given
luminosity compared to observation, if we chose an appropriate ratio
of envelope mass to total stellar mass. The radius of the 3D models is however
larger for a given period compared to previously found period-radius
relations from 1D pulsation models.
The reason for this is not entirely clear,
as the 3D models might appear larger owing to a more extended atmosphere
inflated by the dynamical pressure of small-scale shocks,
or the difference in the representation of the interior
(description of convective energy flux,
dynamical pressure in the interior,
treatment of the stellar core, etc.)
might play a role.
Higher gravity models have a clearly defined pulsation period,
whereas lower gravity objects show a much more irregular behavior,
depending on the relative size of the convection cells
and the typical convective flow speed.

The convective cells themselves do not reach out into visible layers.
However, the network of shocks propagating into the (partially convectively unstable)
atmosphere gives rise to short-lived spatial inhomogeneities across the
stellar disk, which might be the cause for observed dynamical features.

In the convection zone, the dynamical pressure
already reaches significant values
($P_\mathrm{dyn}/P_\mathrm{gas}$\,$\sim$\,5\,--\,20\,\%),
which have to be accounted for in detailed stellar structure models.
However, in the atmosphere,
the dynamical pressure exceeds the gas pressure (by a factor of up to 10).
Both global shocks induced by radial pulsations
and small-scale shocks (not accounted for in 1D models) contribute
to the levitation of material
that can give rise to a ``MOLsphere''
and that allows dust to form.

Comparisons of the properties of the pulsation found in the 3D models
with those from 1D models and observations
give us confidence that the 3D models provide a reliably qualitative
description of the outer convection zone and
the inner part of the atmosphere of AGB stars.
However, further work is needed
(e.g., using non-gray
radiative transfer,
a grid with a larger range of stellar parameters,
a larger computational box,
and a description of dust)
to make the 3D models ready for a more detailed quantitative confrontation with observations.

%#########################################################################################
\begin{acknowledgements}

This work has been supported by the Swedish Research Council (Vetenskapsr{\aa}det).
The computations were performed on resources (``milou'') provided by SNIC through
Uppsala Multidisciplinary Center for Advanced Computational Science
(UPPMAX) under Project p2013234.
We thank Kjell Eriksson for his helpful comments on the manuscript.
\end{acknowledgements}

%#########################################################################################
\bibliographystyle{aa}    % style aa.bst
\bibliography{aa_redsg}

\end{document}